\documentclass[fleqn,10pt]{wlscirep}
\usepackage[utf8]{inputenc}
\usepackage[T1]{fontenc}
\usepackage{tikz}
\usetikzlibrary{shapes,arrows,automata}
\usepackage{dot2texi}
\usepackage{caption}
\usepackage{subcaption}
\usepackage{epstopdf}
\AppendGraphicsExtensions{.tif}
\newcommand{\etal}{{\em et al.}}
\long\def\omitit#1{}
\title{Assessing individual risk and the latent transmission of COVID-19 in a population with an interaction-driven temporal model}

\author[1,+]{Yanir Marmor}
\author[1,+]{Alex Abbey}
\author[2]{Yuval Shahar}
\author[1,*]{Osnat Mokryn}
\affil[1]{Information Systems, University of Haifa, Haifa, Israel}
\affil[2]{Software and Information Systems Engineering, Ben Gurion University, Beer Sheva, Israel}

\affil[*]{omokryn@is.haifa.ac.il}

\affil[+]{these authors contributed equally to this work}

\begin{abstract}
Interaction-driven modeling of diseases over real-world contact data has been shown to promote the understanding of the spread of diseases in communities. This temporal modeling follows the path-preserving order and timing of the contacts, which are essential for accurate modeling. Yet, other important aspects were overlooked. Various airborne pathogens differ in the duration of exposure needed for infection.   Also, from the individual perspective, Covid-19 progression differs between individuals, and its severity is statistically correlated with age. Here, we enrich an interaction-driven model of Covid-19 and similar airborne viral diseases with (a) meetings duration and (b) personal disease progression. The enriched model enables predicting outcomes at both the population and the individual levels. It further allows predicting individual risk of engaging in social interactions as a function of the virus characteristics and its prevalence in the population. We further showed that the enigmatic nature of asymptomatic transmission stems from the latent effect of the network density on this transmission and that asymptomatic transmission has a substantial impact only in sparse communities.
\end{abstract}
\begin{document}

\flushbottom
\maketitle
%
%
\thispagestyle{empty}

\section*{Introduction}
The SARS-CoV-2 pandemic, like other diseases, spread differently in different countries and communities~\cite{dufresne2020r0}. Disease progression results from the interplay between the population's complex interaction dynamics, which are associated with the population’s physical contacts’ network, and the disease dynamics~\cite{sayama2015introduction,liu2018measurability}, as well as characteristics such as the population age~\cite{inde2021age,banholzer2022estimating}. 

 Here, we present in detail an SEIR-like {\em interactions-based contagion model} (ICM) of airborne diseases for COVID-19~\cite{abbey2022interaction,abbey2022analysis} over real-world interaction data that is enriched with personal disease progression details that depend on the individual \textit{susceptibility} to the disease. The model is termed {\em Interactions-based Contagion Model with Individual outcomes} (ICMI). The individual susceptibility determines the severity of the disease for an infected individual. For COVID-19, the probability of an individual contracting a severe form of the disease, referred to as their susceptibility to the disease, is a function of a person's age~\cite{inde2021age}. 

 Accurately modeling disease progression requires considering the time-respecting paths, which are the sequence and order of these interactions between the members~\cite{holme2012temporal,ENRIGHT201888,holme2021fast}. Contacts' temporal ordering and dynamics are crucial for understanding the transmission of infectious diseases. The interactions' temporal path ordering was shown to affect the spreading dynamics~\cite{rocha2011simulated,scholtes2014causality,holme2015information,delvenne2015diffusion,grossmann2020importance,masuda2020small,wang2021impact}. Further, considering the accurate structure of human interactions is pivotal for correctly predicting the spread of epidemics, such as the Covid-19 disease~\cite{herrmann2020covid}.
To model the disease progression in a real-life community, we use the real-world encounters data from the Copenhagen Networks Study (CNS) dataset~\cite{stopczynski2014measuring,Stopczynski:2015aa,sapiezynski2019interaction}. The use of real-world interaction data as the dataset for our analysis allows for a person-to-person spread of disease at the level at which it occurs in its actual temporal and local contexts~\cite{vespignani2020modelling,thurner2020network}.

 An investigation of the transmission of a severe acute respiratory syndrome (SARS) in a 2003 Toronto-area community outbreak found that ``longer and closer proximity exposures incurred the highest rate of disease''~\cite{rea2007duration}.  Thus, the duration of the interaction is crucial for correctly simulating the airborne transmission of various pathogens. 
 A key feature of ICM is that the interaction's duration is correlated positively with the latent transmitted viral load if the encounter is with an infected person.
 Duration of interaction as a proxy for the transmitted viral load was used for estimating viral infection between animals~\cite{wilber2022model} and  humans~\cite{smieszek2009mechanistic,muller2020mobility,mo2021modeling,nagel2021realistic,lorch2020quantifying}. 
Smieszek~\cite{smieszek2009mechanistic} noted that the time people spend interacting with each contact person decreases as the number of contact persons increases. The suggested model takes the length of the exposure into account while following real-world interactions of various scales that exist in the CNS dataset, as the proximity information enables the detection of gatherings~\cite{sekara2016fundamental,ciaperoni2020relevance}.


%

%

Combining the above with individual susceptibility, the ICMI model presents three significant contributions. 

First, it  extends the capabilities of ICM~\cite{abbey2022analysis,abbey2022interaction} by taking individual outcomes into account, enabling it to predict outcomes at the community and individual levels. With ICMI, the daily and average expected percentage of needed hospital beds could be predicted for communities that leverage digital proximity tracing applications~\cite{cencetti2021digital,pung2022using}.  In addition, the model considers the probability of symptomatic or severe infection in correlation with the individual's age. It models age using a personal susceptibility parameter, $s_i$, a normalized parameter denoting the personal susceptibility to the disease that correlates with age. Thus, lower values of $s_i$ correspond to younger people who are more likely to be asymptomatic. 

Second, ICMI enables us to model and predict individual risk of infection given personal daily exposure. Individual risk prediction projections are provided as a function of different variants' virality and individual vulnerability probability due to personal immune levels. 
This prediction of individual risk not only enables individuals to plan their schedules to reduce their risk of infection but also presents a complementary paradigm to the current government-imposed non-pharmaceutical interventions (NPIs), providing an additional layer of personal control.

Third, the ICMI model contributes to the study of the spread of diseases caused by asymptomatic transmission.
Asymptomatic transmission of diseases such as COVID-19 is considered to be the Achilles' heel of the pandemic control~\cite{gandhi2020asymptomatic,shahar2021computing} since asymptomatic individuals continue their normal social and travel activities while being hard to trace~\cite{PARK2020100392}.
The exact extent and impact of asymptomatic transmission are debatable~\cite{byambasuren2020estimating}, and differences in asymptomatic cases' generation time may affect the transmission and spreading factor estimations~\cite{park2020time}. Here, we find that asymptomatic transmission is significant and influential only in relatively sparse networks, and its effect is mitigated in dense networks. 
Asymptomatic transmission influences the progression of the disease only in sparse communities.

Thus, taking into account the macroscopic daily outcomes of the microscopic interactions while considering individual susceptibility to severe disease enables us to predict outcomes at both the population and the personal level. The model further allows predicting individual risk of conducting meetings as a function of the virus characteristics and its prevalence in the population. We further showed that the enigmatic nature of asymptomatic transmission stems from the latent effect of the network density on this transmission and that asymptomatic transmission has a substantial effect only in sparse communities.

\section*{Results}
\begin{figure}[!ht]
\centering
    \includegraphics[width=.7\textwidth]{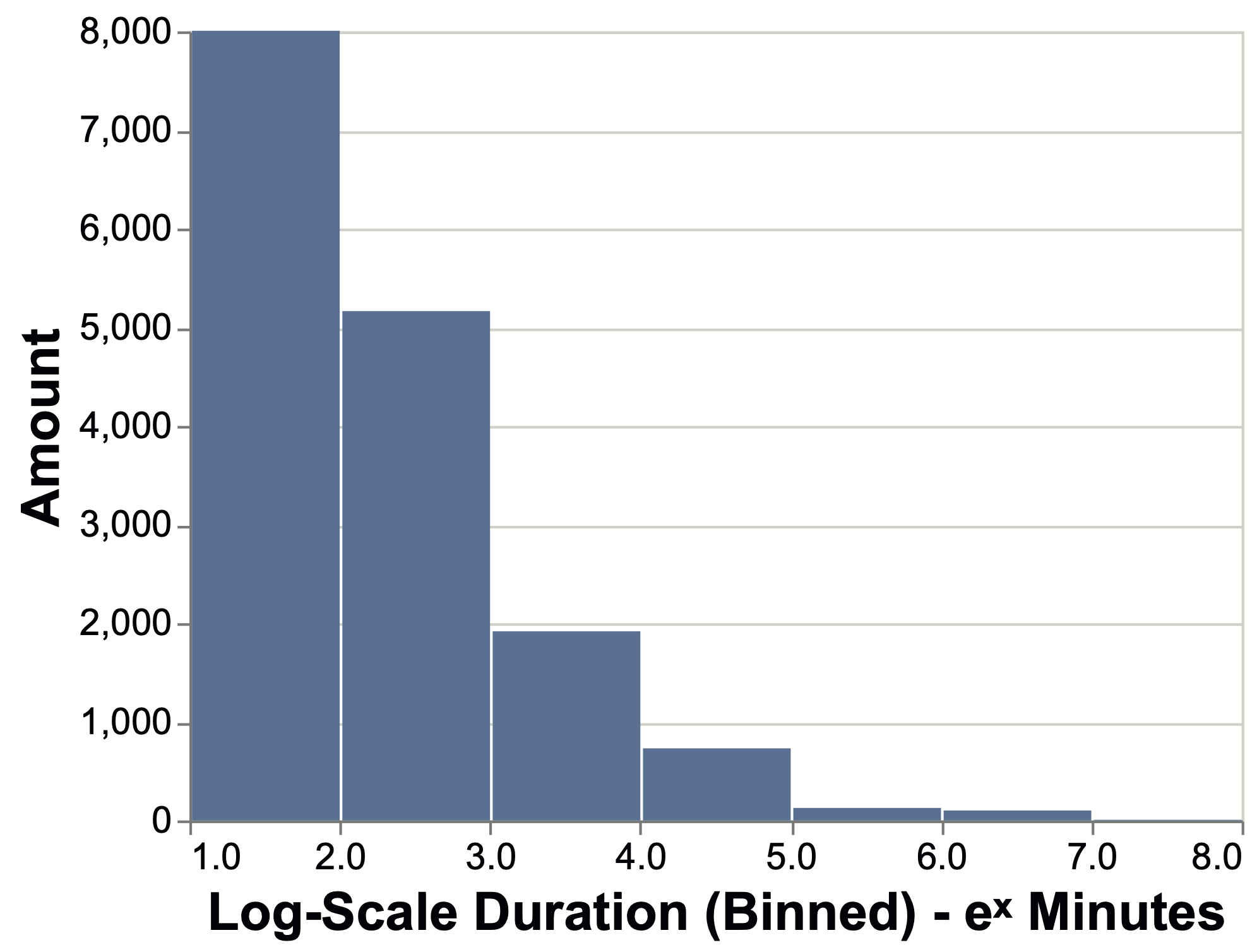} 
\caption{The average daily meetings duration histogram of the CNS social networks. X-axis depicts a log-scale measure of meetings' duration,  the y-axis denotes the average daily amount of meetings of various duration in the CNS network measured over 24-hour intervals.}
\label{fig:td}
\end{figure}
We ran numeric simulations over the CNS real-world interaction dataset, combining two processes. The first is the interaction model (ICM). The model considers the duration of the interactions, as is described in Equation~\ref{eq:2}. Figure~\ref{fig:td} shows the average daily meetings' duration in the CNS dataset. Considering the dataset consists of close to 700 individuals connected daily, the data demonstrate a skewed distribution graph, with many short meetings and a few very long ones. The model then accounts for the circadian nature of human behavior by considering the daily probability of not being infected in any of the encounters (Equation~\ref{eq:vl2}).

The second process incorporated is the personal disease progression modeling according to the individual susceptibility, as described in the infection model presented in Fig \ref{fig:timeline}.  The CNS dataset was recorded in a university and contained the reading of 700 students. The interactions are probably denser than that of a small metropolitan~\cite{abbey2022interaction}. The ICMI model, however, assigns heterogeneous personal vulnerability values as described in each experiment.  Personal vulnerability correlates with age, as $s_i$ is a normalized parameter that correlates with age, $s_i \in [0 \ldots 1]$.   Low values of $s_i$ denote very young people (kids), and high values denote very old.  

The two processes are combined in each of the simulations, and each experiment is the result of $200$ iterations. The simulation code is written in Python and is freely available: \url{https://github.com/ScanLab-ossi/covid-simulation}.

\subsection*{COVID-19 disease progression with individual outcomes}
\begin{figure}[h]%
\centering
 \begin{subfigure}[b]{\linewidth}
    \centering
    \includegraphics[width=0.95\linewidth]{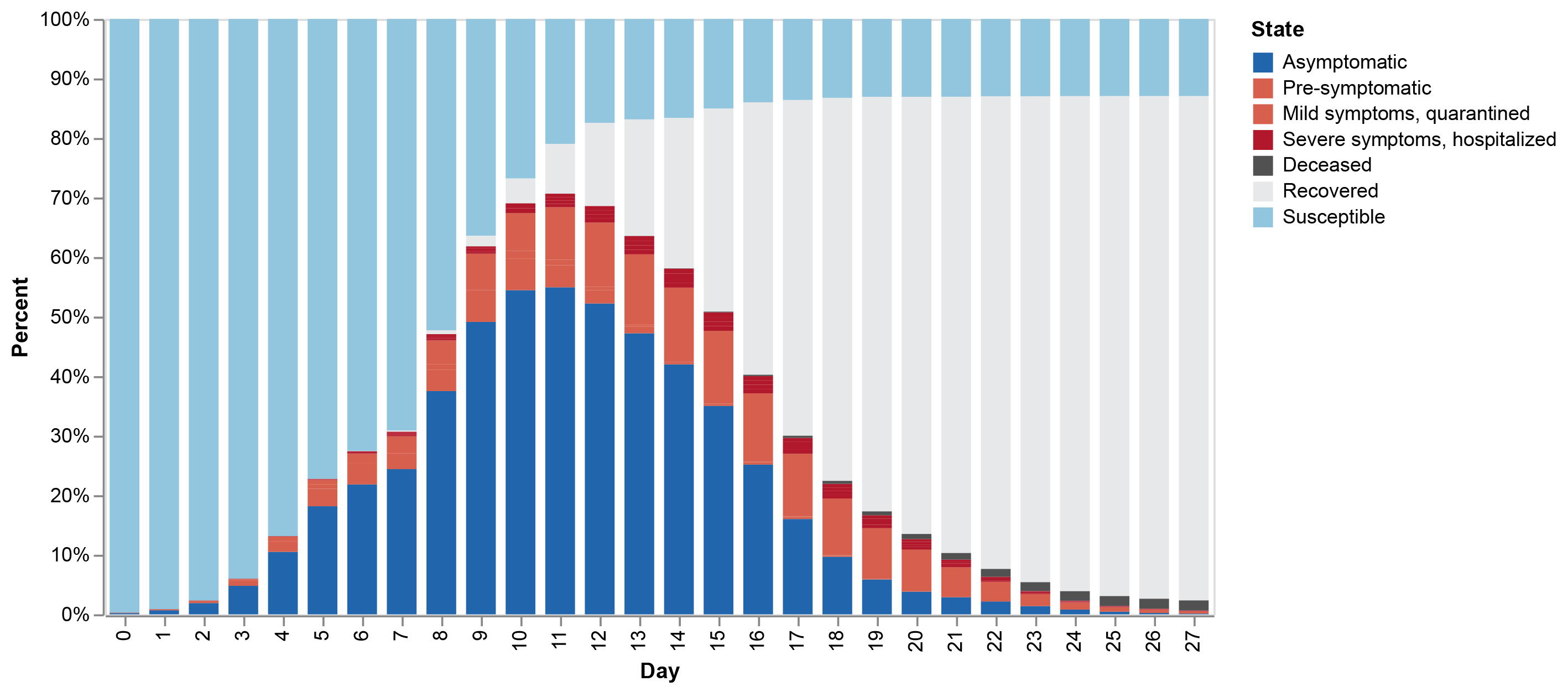} 
    \caption{Disease progression - Fast variant} 
    \label{fig:icmi-5} 
  \end{subfigure}
  \begin{subfigure}[b]{\linewidth}
    \centering
    \includegraphics[width=0.95\linewidth]{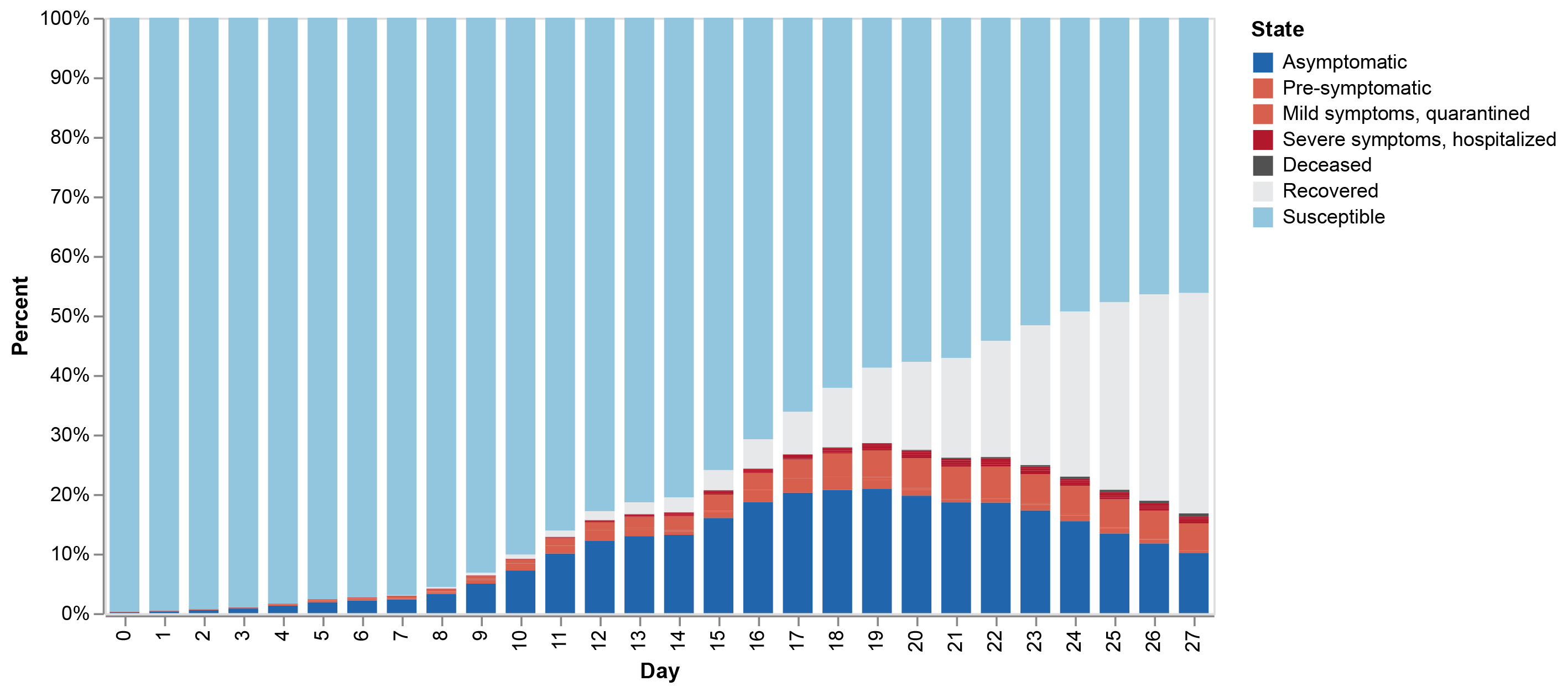} 
    \caption{Disease progression - Slow variant} 
    \label{fig:icmi-60} 
  \end{subfigure} 
\caption{COVID-19 Disease progression with individual outcomes at the community level for  the real-world CNS network: (a) Fast variant, s.t. every encounter has the potential to infect ($D_{\text{min}}$ is minimal); (b) Slow variant, s.t. only encounters that are at least 60 seconds have the potential to infect ($D_{\text{min}} > 60$)}
\label{fig:icmi}
\end{figure}

We start with a simulation of the disease progression over the real-world CNS temporal network. Given a variant with a minimum exposure latency $D_{\text{min}}$ (i.e., each encounter with an infectious node is long enough to infect), we compute disease progression as well as individual outcomes in a community. Here, we start with one infectious initial node, i.e., a single patient zero. $P_{\text{max}}$, the maximal probability of getting infected given exposure is set to $0.5$, and the population is young ($s_i$ values are small), s.t. 80\% of the population is asymptomatic (All parameters are configurable). 

During each day, for each node $i$, interactions with {\em Infectious} nodes, either presymptomatic or asymptomatic, that are longer than $D_{\text{min}}$ are examined. At the end of each day, according to Eq.~\ref{eq:2} and Eq.~\ref{eq:vl2} a node either stays in state $S$usceptible or enters state $E$xposed and infected. Personal progression of the disease then follows the individual state machine depicted in Figure~\ref{fig:timeline}.
Figure~\ref{fig:icmi} depicts the disease progression over the CNS real-world temporal network for two different COVID-19 variants. Each experiment was the result of $200$ iterations. Figure~\ref{fig:icmi-5} shows the progression of a fast variant ((that is, $D_{\text{min}}$ is low), for which every encounter with an infectious individual is long enough to infect, and Figure~\ref{fig:icmi-60} shows the disease progression and community outcome for a slower variant, in which only long exposures can infect (that is, $D_{\text{min}}$ is high). 
Individual outcomes can predict the number of daily hospital beds needed, denoted in red in the figures, and the expected death toll, if any.

We further see that more contagious variants, as is the case shown in Figure~\ref{fig:icmi-5}, create, even in young populations in which the majority of patients are asymptomatic, a higher load on the community hospitals.

\subsection*{Individual projection of infection as a function of the daily exposure}
Considering the changing daily number of meetings and their lengths, we can approximate the individual probability of infection, given their daily exposure. We project here individual infection outcomes as a function of the daily number of meetings and their lengths and contextual information. Contextual information can be considered a proxy for the virality of a variant and susceptibility to the infection of an individual's immune system.

 When considering meetings' duration, the minimum duration in the model, $D_{\text{min}}$, is used as an approximate for the virality of a variant, with very low $D_{\text{min}}$ values corresponding to very infectious variants, and vice-versa. The other global parameter in the model, $P_{\text{max}}$, is used here on a per-individual level, $P_{{\text{max}},i}$, to denote the personal vulnerability of a person $i$ to the virus. In this approximation, high $P_{{\text{max}},i}$ values correspond to highly vulnerable, e.g., immune-compromised people. Lower values correspond to a lower probability of getting infected, given maximal exposure. In both experiments, the probability of getting infected is computed. The severity of the infection would then depend on the individual susceptibility.

 In these two experiments, we examine the probability of becoming infected. This probability differs from the probability of being symptomatic or severely ill, which correlates with the personal susceptibility parameter, $s_i$. The probability of getting infected does not correlate with this parameter, of course~\cite{shahar2023statistical}.

\begin{figure}[h]%
\centering
\includegraphics[width=.7\textwidth]{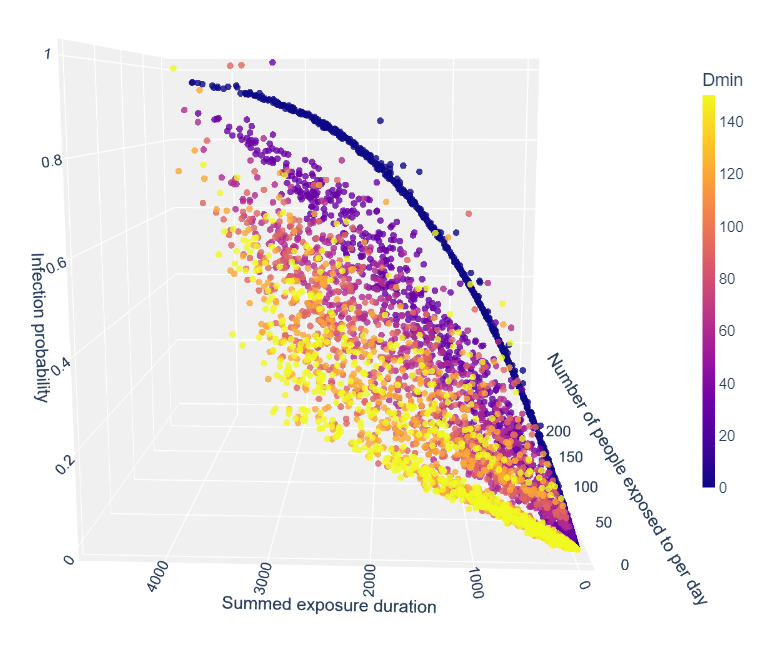}
\caption{Individual probability of an adult for getting infected in an average day as a function of the number of exposures and their aggregated duration, according to the model, for various $D_{\text{min}}$ values. Different $D_{\text{min}}$ values, which denote the minimal duration needed for infection, correspond to different variants, with Lower $D_{\text{min}}$ values corresponding to more infectious variants, i.e., faster pathogens.}\label{fig:pod}
\end{figure}

\begin{figure}[h]%
\centering
\includegraphics[width=.7\textwidth]{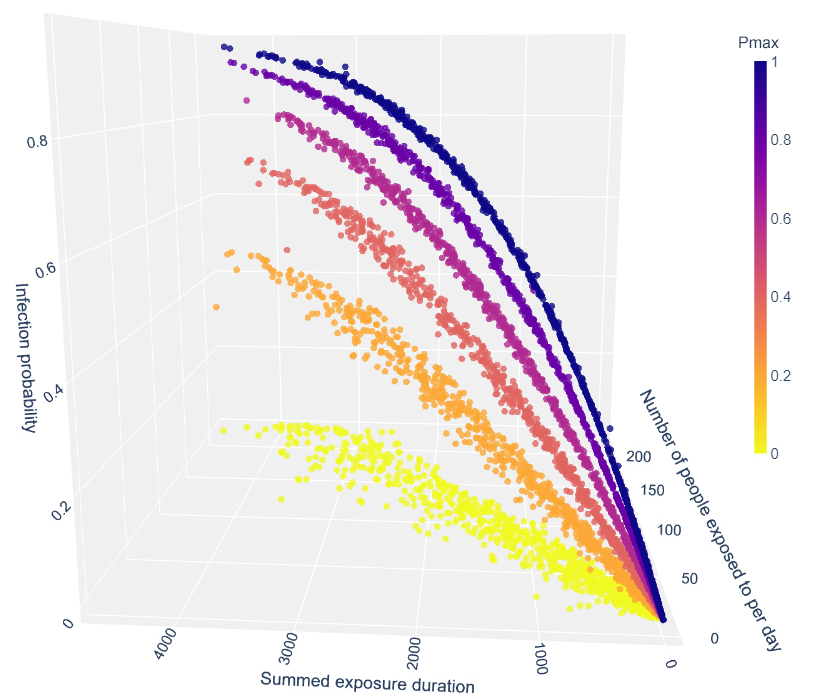}
\caption{Individual probability of getting infected in an average day as a function of the number of exposures and their aggregated duration, according to the model, for various $P_{\text{max}}$ values. Different $P_{\text{max}}$ values correspond to individual  vulnerability probability, depending on personal immunization and inoculation  status.}\label{fig:pop}
\end{figure}

We generate random exposure data sampled from distributions fitted to the CNS real-world aggregated data. Aggregated information was taken such that an average daily outcome could be calculated. The average number of infected individuals differs on different days, depending on the spread of the virus in the community. Typically, a person would not be aware of the precise situation in their community. Hence, we consider the average of all the days in the dataset as the typical day over which we calculate individual outcomes.  
In a time window $\tau$, the information for the possible $N$ people encountered by an average person during that time window was generated by randomly sampling from a distribution $N \sim ax^b$  fitted to the distribution of the CNS dataset ($a\approx 0.051, b\approx	-0.635$, $max\{\mathrm{Cov}[N]\}\leq4.4e^{-5}$). Meetings lengths, ${d_{ik}^j}$, were generated from a discrete probability distribution based directly on the distribution of all encounters duration in the original dataset. Each experiment is the result of $200$ iterations, and the results are depicted in Figures~\ref{fig:pop} and~\ref{fig:pod}, showing the personal projection of infection as a function of the daily exposure. Understandably, the number of contacts an individual is exposed to per day correlates highly with that individual's total exposure per day.
The most basic configuration ($D_{\text{min}}=0$ and $P_{\text{max}}=1$) for an individual shows a close to a linear relationship between the daily amount of nodes exposed to, the sum of exposure duration and $P_{\text{infected}}$.

Figure~\ref{fig:pod} depicts the projected individual outcome for various levels of exposure to variants, approximated using the amount of viral load needed for infection. Higher  $D_{\text{min}}$ values correspond to a higher viral load needed for infection and, thus, to a less transmissible variant. Lower $D_{\text{min}}$ values correspond to a highly transmissible variant. More encounters, even if short, increase the risk of infection even when the virus is less infectious. However, individuals can lower their risk of infection by lowering their daily exposure to less transmissible viruses.

Figure~\ref{fig:pop} depicts the projected infection probability as a function of the daily exposure for various individual vulnerability levels. Here, the most transmissible variant was considered, i.e., enough viral load is transmitted even in very short exposures. Here, recovered or vaccinated people have a significantly lower probability of infection given similar exposure. Lowering $P_{\text{max}}$ reduces $P_{\text{infected}}$ by that factor, with the slope of the linear correlation dropping accordingly.  

\subsection*{Temporal density mitigates the effect of asymptomatic infection}
Modeling asymptomatic transmission can be achieved using the individual susceptibility parameter at the community level. 
ICMI's detailed individual disease progression encompasses the individual susceptibility factor per node, $s_i$. At the community level, we define $\vec{S_c}$ as the vector of personal susceptibility levels in a community. $\vec{S_c}$ is defined as the vector $s_i, i \in [1..n]$, with $n$ is the total number of individuals in the simulation. The distribution of community personal susceptibility levels, $\vec{S_c}$, determines the percentage of symptomatic and asymptomatic individuals in the simulation. Thus, it can be used to examine the effect of different levels of asymptomatic carriers in the population on the progress of the disease. As susceptibility correlates with age, younger populations will yield more asymptomatic patients. 

\begin{figure}[h]%
\centering
    \includegraphics[width=0.4\linewidth]{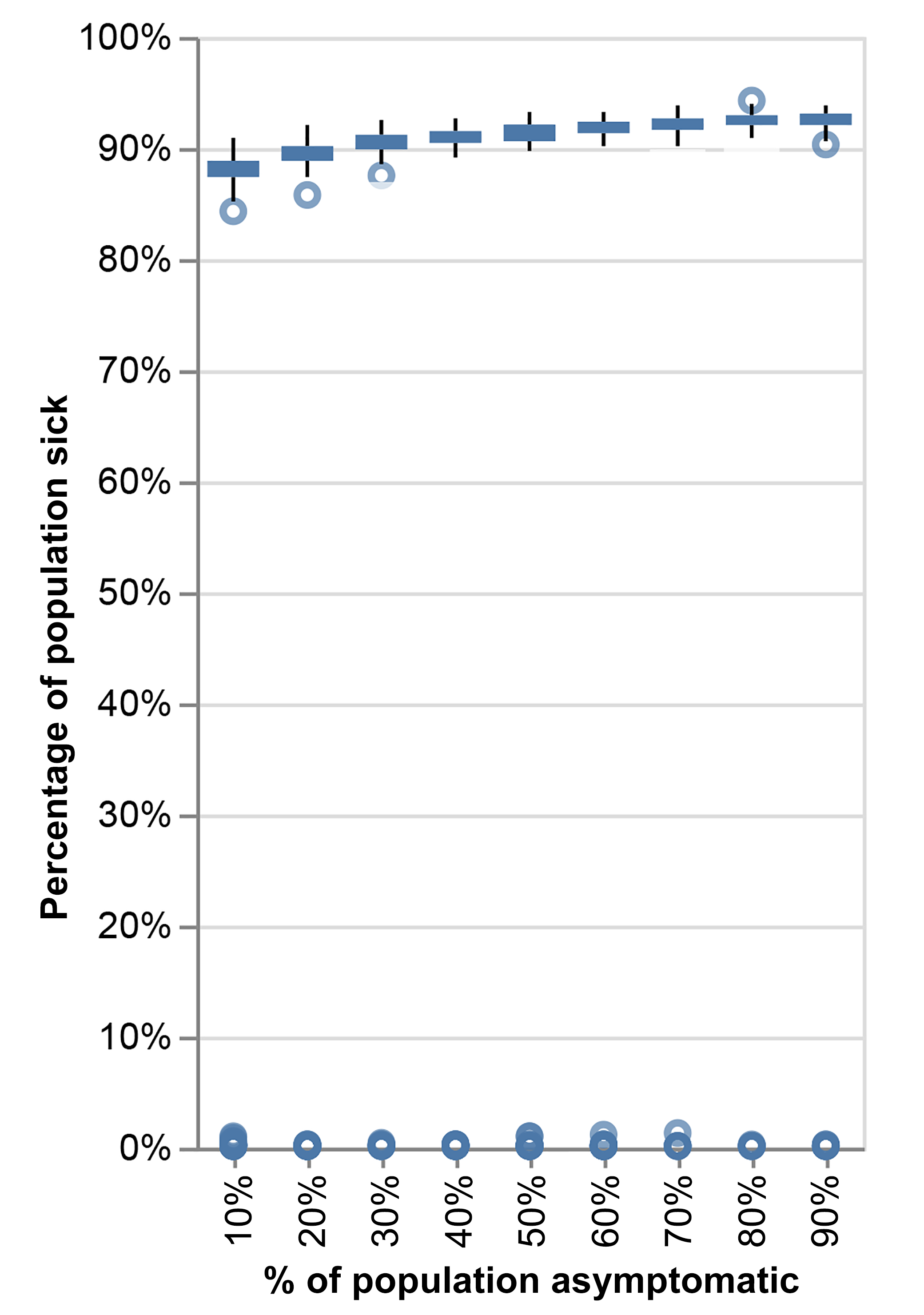} 
\caption{Quantifying the effect of asymptomatic transmission. The younger population is more inclined to be asymptomatic, as captured by the susceptibility parameter in the model. The x-axis denotes the percentage of the young population in the experiment, modeled as asymptomatic, and the y-axis denotes the rate of infection in the population.  Each experiment was performed 200 times, with random placement of the initial patient zero. }\label{fig:as-orig}
\end{figure}
To examine the effect of asymptomatic transmission in a population, we ran the simulation over the CNS dataset while varying the average age of the population, s.t., the percentage of asymptomatic among the infected varies between 10\% to 90\%. Each experiment was performed 200 times, with random placement of the initial patient zero. Figure~\ref{fig:as-orig} depicts the results of the experiment. Surprisingly, we find that asymptomatic transmission does not have a significant effect in this experiment. This result aligns with the difficulty of assessing the true impact of asymptomatic transmission~\cite{subramanian2021quantifying}. 

This is also a sensitivity test for how transmission in a community differs when the population age varies. 

\omitit{
\begin{figure}
\centering
 \begin{subfigure}[t]{0.3\textwidth}
    \includegraphics[width=\textwidth]{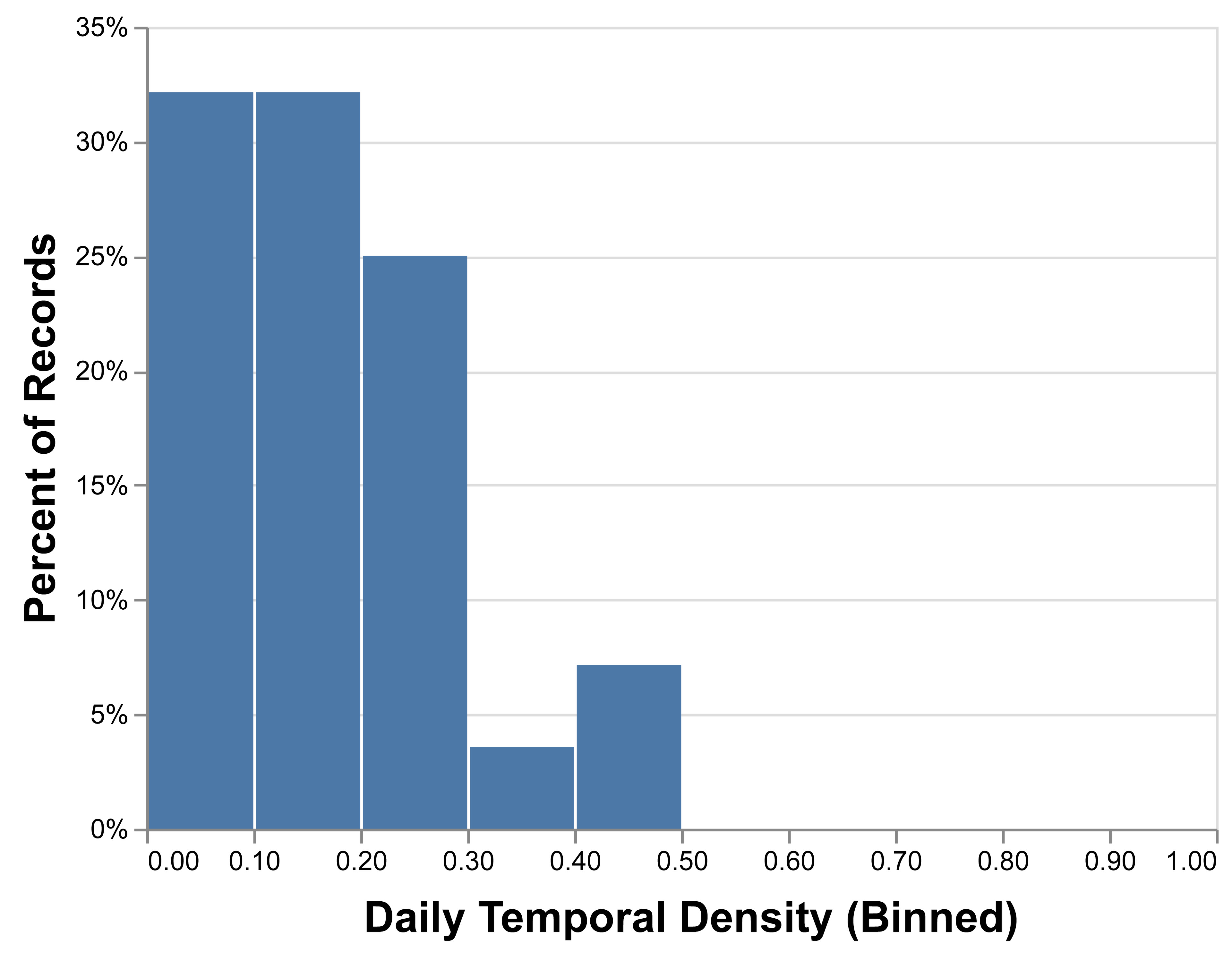} 
    \caption{CNS network} 
    \label{fig:td1} 
  \end{subfigure}
   \begin{subfigure}[t]{0.3\textwidth}
    \includegraphics[width=\textwidth]{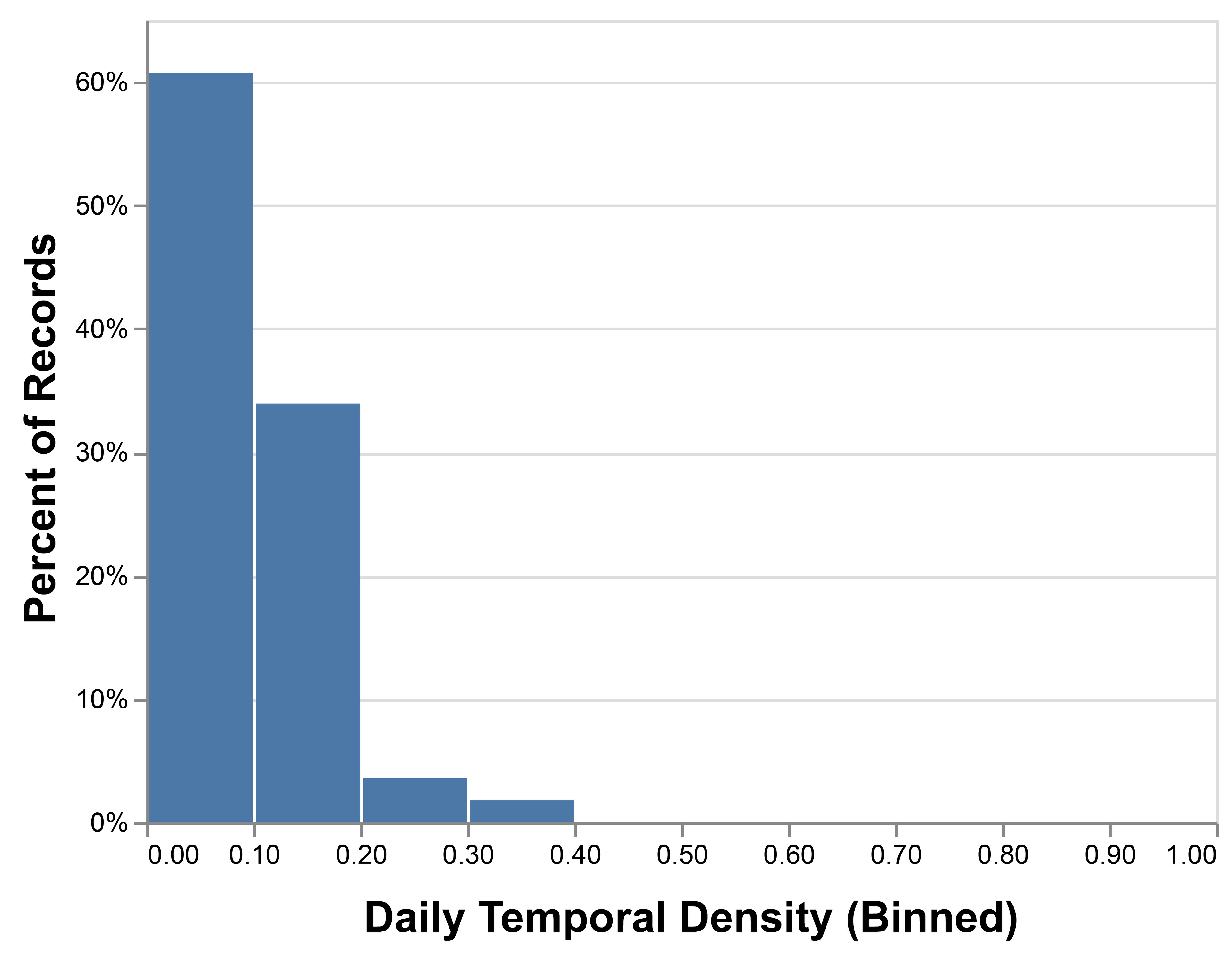} 
    \caption{Half density} 
    \label{fig:td2} 
  \end{subfigure}
   \begin{subfigure}[t]{0.3\textwidth}
    \includegraphics[width=\textwidth]{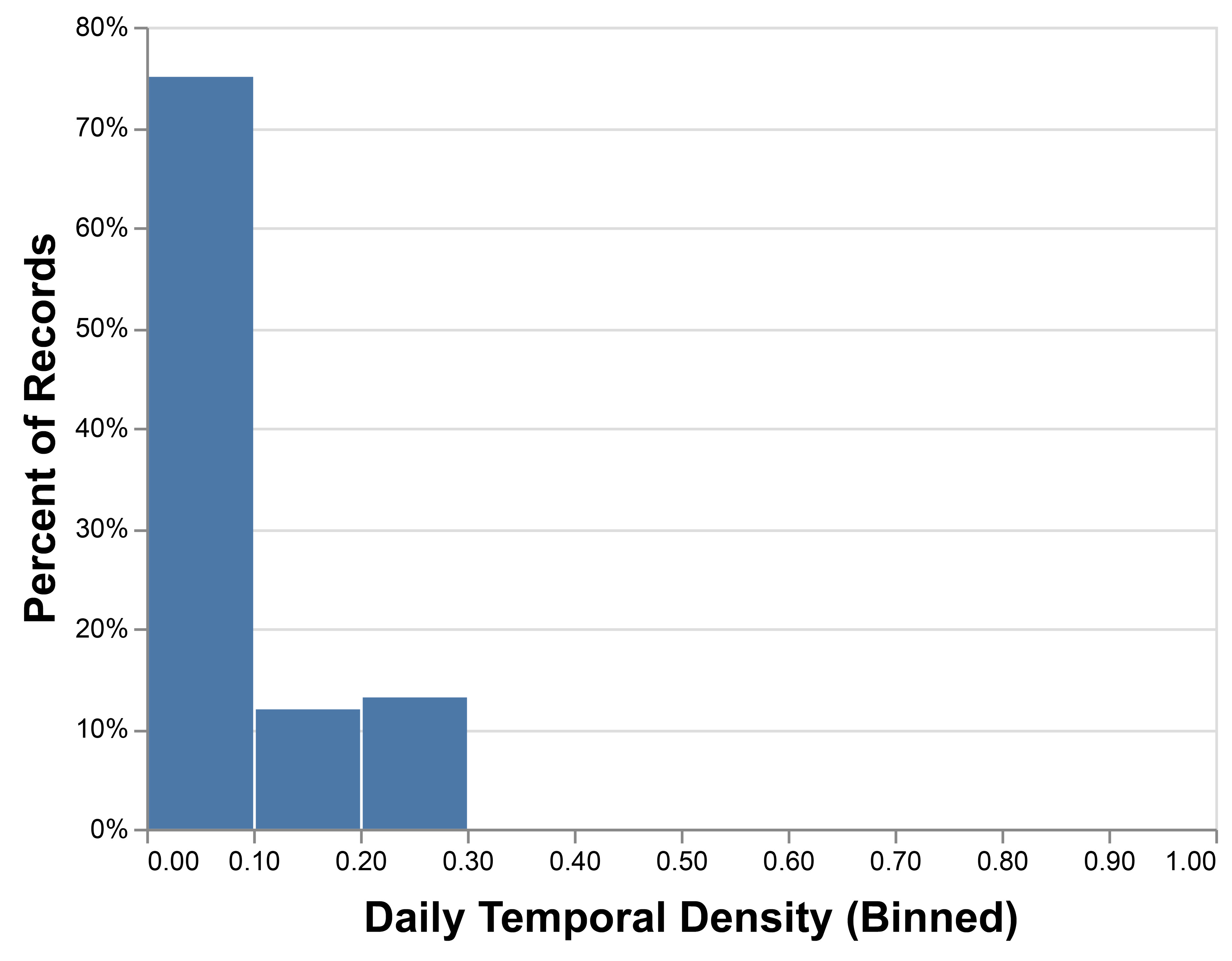} 
    \caption{Third density} 
    \label{fig:td3} 
  \end{subfigure}
\caption{Daily temporal density histogram of the CNS social networks. X-axis counts the number of days with the same temporal density, y-axis denotes the temporal density percentage: (a) Original temporal network; (b) Reduced density - CNS with half the original temporal density and twice days; (c) Reduced density - CNS with third the original temporal density and therefore three times the number of days}
\label{fig:cnsd}
\end{figure}
}
In a previous study~\cite{abbey2022interaction}, we demonstrated that the CNS network is very dense.  
Over two-thirds ($\sim 64\%$) of the days have a temporal density of $0.2$ or lower. That is, the overall number of interactions is at most 20\% of the possible number of interactions. A quarter of the days have a temporal density of 25\%, and the remaining 11\% days are very dense, up to half the maximal possible density in which everybody meets everybody. 

We perform the following experiments to understand the effect of temporal density on latent transmission due to asymptomatic people. We reduce temporal density by splitting each day to $k=[2,3]$ pseudo days with $1/k$ the interactions of the original day. Thus, for example, a network with half the density will be depicted as twice as long, pseudo-days wise. We then repeat the experiment described above, changing the percentage of asymptomatic in the population over the longer, less dense networks. 

\begin{figure}[h]%
\centering
  \begin{subfigure}[b]{0.46\linewidth}
    \centering
    \includegraphics[width=0.95\linewidth]{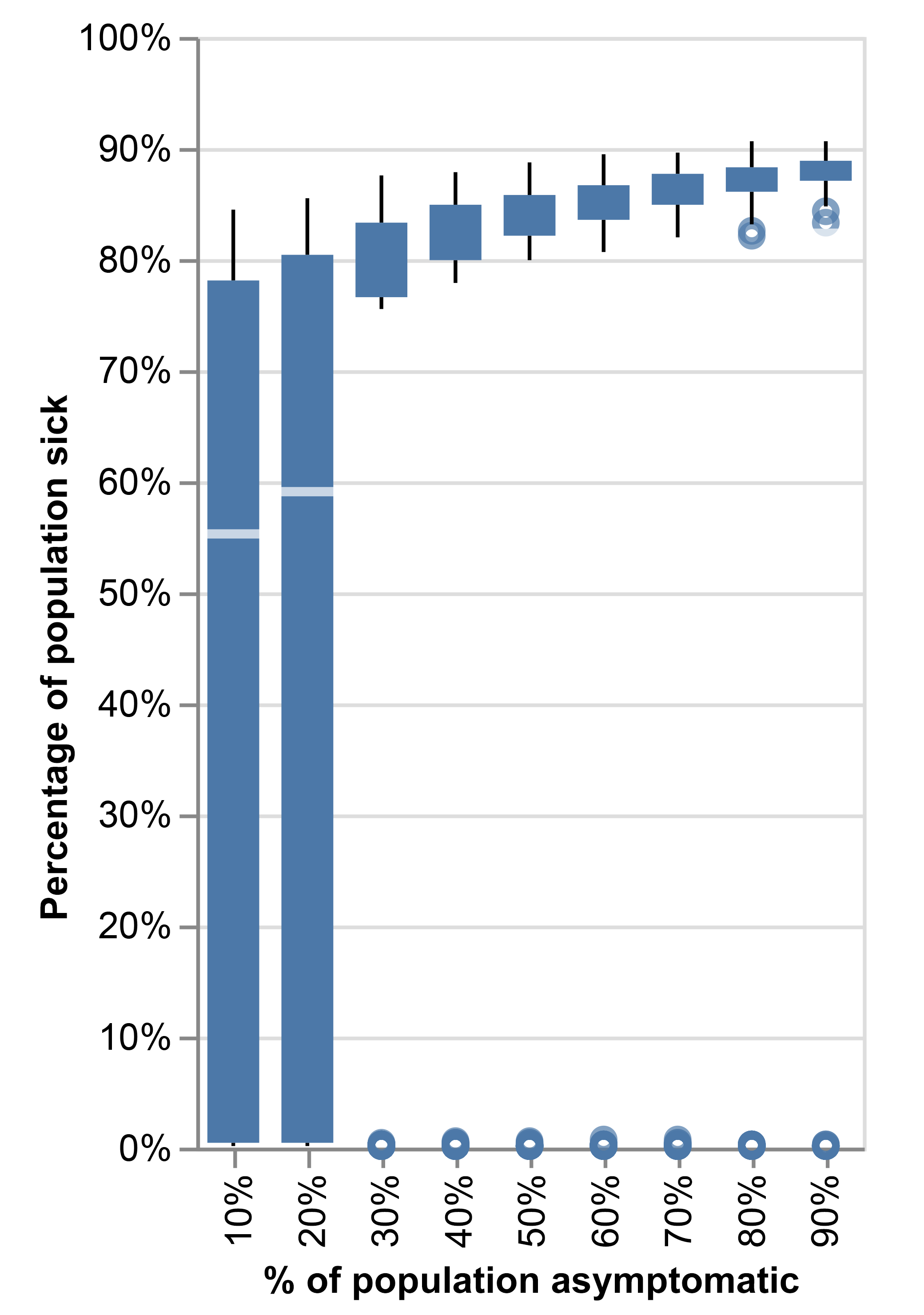} 
    \caption{Half temp. density} 
    \label{fig:ab} 
  \end{subfigure} 
  \begin{subfigure}[b]{0.46\linewidth}
    \centering
    \includegraphics[width=0.95\linewidth]{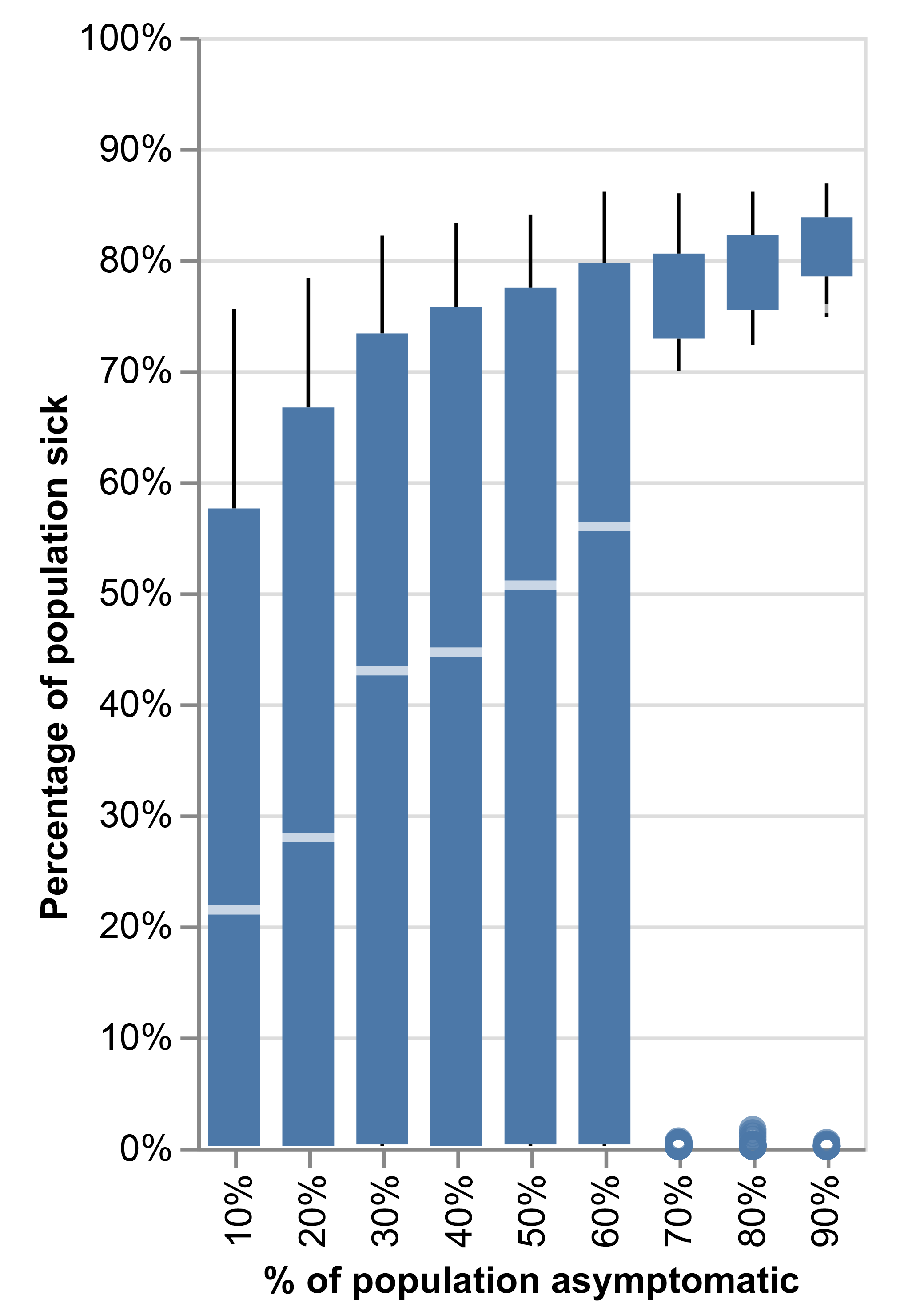} 
    \caption{One-third density} 
    \label{fig:ac} 
  \end{subfigure}
\caption{The effect of latent transmission due to asymptomatic infections on the spreading of the disease with different temporal density conditions in the real-world CNS network: (a) Reduced density - network with half the original temporal density; (b) Reduced density - network with one-third the original temporal density }\label{fig:abc}
\end{figure}
Figure~\ref{fig:abc} demonstrates the effect of temporal density on the latent transmissibility of a disease. In each experiment, we changed the percentage of asymptomatic carriers in the population between 10\% and 90\%. Each experiment was performed 200 times, with random placement of the initial patient zero. Figure~\ref{fig:ab} depicts the spread of the disease on a network with half the temporal density of the original one, showing a significant effect of the latent, asymptomatic transmission. Similarly, when the temporal density is one-third that of the original network, as is the case in Figure~\ref{fig:ac}, the more asymptomatic the population, the more they contribute to the total infection rate. 

Hence, we find here that asymptomatic transmission influences the total infection rate in sparse networks. However, increased density mitigates the effect of asymptomatic transmission, and in highly dense communities, we could expect the disease to spread fast, regardless of age distribution.

\section*{Discussion}
\label{sec4}
Here, we presented the ICMI model, which assesses the disease progression over real-world community interactions. The model follows the temporal dynamics of the interactions while taking into account the virus parameters and individual susceptibility to the disease. Three different aspects combine the model: following temporal interactions, considering the interaction duration as a proxy for the transmitted viral load, and incorporating individual susceptibility. We discuss here each of these aspects.  

The last decade's abundance of temporal information paved the path to further understanding of the temporal dynamics of networks~\cite{lazer2009life,holme2012temporal,mokryn2016role}.
Temporal networks have become the playground for inferring behavior in a plethora of areas, including, but not limited to, the inference of the effects of changes on the evolution of networks~\cite{Peel2015,miller2020size}; revealing hidden structures, such as the revelation of the structure of ``co-presence'' in a metropolitan or university from temporal daily encounters~\cite{sun2013understanding};   the temporal, spatial diffusion of information in social media~\cite{si2020comparative}, and the behavior of viral processes over it~\cite{blondel2015survey}.
Incorporating temporal dynamics into existing viral models has proven challenging. Recently, Holme~\cite{holme2021fast} suggested a fast implementation of a temporal SIR over temporal networks. The model, however, does not allow for dynamics representative of real-world interactions, e.g., it assumes exponential time to recovery. Accounting for large infection events, Cooper \etal~\cite{cooper2020sir} extended the SIR model to consider the surges in the size of the susceptible population over time.  Yet, they do not consider other population dynamics. 
Recent agent-based models incorporated human dynamics such as inter-arrival times and heterogeneity of interactions~\cite{kerr2021covasim,truszkowska2021high}.
For example, OpenABM~\cite{hinch2021openabm} created a simulated city environment of 1 million people and their dynamics, typical of a household, schools, social interactions, etc. Their agent-based modeling was then used to evaluate different social distancing techniques in that environment. Unlike the presented ICMI model, they do not consider personal differences, i.e., individual susceptibility, yet discuss their importance for accurate modeling. 
Agent-based models, however, do not maintain other real-world dynamics such as temporal path ordering.



Several recent models considered that the transmitted viral load can differ between interactions. Given an estimated load, the rate of the shedding of viral load in metropolitan transportation was estimated~\cite{muller2020realistic}. Viral load and meeting duration were considered to understand the interplay between biological and social factors in the asymptomatic spread of the disease~\cite{tadic2021microscopic}. 

The importance of considering personal differences was discussed by many. For example, age and comorbidity factors contributed to the appearance of symptoms and to early isolation after infection~\cite{grossmann2021heterogeneity,gnanvi2021reliability}. Specifically, the SIDARTHE model~\cite{giordano2020modelling}  differentiated between symptomatic and asymptomatic patients and considered the severity of their symptoms as a proxy for isolation.  SIDARTHE used the following states: susceptible (S), infected (I), diagnosed (D), ailing (A), recognized (R), threatened (T), healed (H), and extinct (E). However, the model did not consider the temporal dynamics of human interactions nor the effect of the duration of the encounters.
Our model agrees with the findings of Peirlink \etal \cite{peirlinck2020visualizing} that if infectiousness is the same for both symptomatic and asymptomatic patients, the size of the asymptomatic population does not largely affect the overall outbreak dynamics. Other complementary findings to ours are those of Park \etal\cite{park2020time} and Subramanian \etal\cite{subramanian2021quantifying}, who have shown that a faster asymptomatic transmission rate increases the realized proportion of asymptomatic transmission. Yet these findings did not take into account how asymptomatic contagion is affected by the density of the population. In here, it was shown that  asymptomatic transmission has a substantial impact only in sparse communities. 


ICMI considers the interaction's length as a proxy for the transmitted viral load. It computes the probability of infection at the end of the day.  ICMI aggregates viral load only in the case of higher-order interactions, like group meetings. Otherwise, it does not consider different interactions as having an additive effect, as there is no evidence that viral load ``remains'' in-between interactions~\cite{hebert2020macroscopic}. The CNS dataset was similarly used by Hambridge \etal~\cite{HAMBRIDGE2021325}, who devised a temporal interaction-based SEIR model to assess the effect of various interventions on the CNS data. Unlike ICMI, the work does not consider interactions' length, and assumes that multiple exposures on the same day increase the risk. 

The ICMI model is the first to consider the interaction between the daily macroscopic dynamics and the microscopic interactions to predict the spreading dynamics in a population, given the various outcomes of getting infected within the population. 
Given today's abundance of digital traces, the ICMI model enables policymakers to assess the disease progression using real-world interactions typical within their communities while considering the various pathogens. We have shown that by incorporating the individual outcomes and the community age distribution, policymakers could receive an estimation of the expected number of hospital beds required as the disease progresses in the community.

Devising a method for predicting individual outcomes as a function of daily exposure to a pathogen further gives a personal design tool for individuals to assess their risk in taking meetings of various lengths given the spread of the virus in the community. The method enables policymakers to decide when would be the right time to restrict large gatherings and long meetings and decide on general guidelines for the public. 


The work has several limitations. The CNS dataset is the result of students' interactions in a university. The interactions were denser than social interactions, which may result in a faster-than-reality infection process~\cite{abbey2022interaction}. We thus showed that the effect of asymptomatic latent transmission is negligible in dense networks but not in sparser ones. An additional limitation is that the analysis of latent transmission due to asymptomatic infections does not consider that  symptomatic and asymptomatic people shed different viral loads~\cite{liu2020viral,ferretti2020quantifying}. The option was implemented in the code and is part of our future work. 

\section*{Conclusions}

The paper describes an SEIR-like interaction-driven contagion model of airborne disease for COVID-19 with individual outcomes (ICMI) that depend on age. It shows that ICMI, encompassing daily macroscopic dynamics with the microscopic level of interaction duration, enables outcomes prediction on both the population and the individual levels. It further allows for individual assessment of risk levels in different populations and can be used as a tool by policymakers. Using ICMI, we further showed that the effect of latent transmission due to asymptomatic infections depends highly on the structure and interaction density of the population’s social network and is higher in sparser structures. 

In future work, we intend to use additional datasets obtained from community digital contact tracing or community structure as provided in~\cite{hinch2021openabm} to explore the model further. The data would be augmented with the corresponding population age distribution to enable community-level predictions for real-life communities. Here, age was considered a proxy for personal vulnerability to the disease since we considered the recent COVID-19 pandemic a key example. However, in the case of other pathogens, other individual factors can be used.

\section*{Methods}

\label{sec3}

\subsection*{Interaction-driven contagion SEIR-like model with personal outcomes (ICMI)}
The ICM with Individual outcomes (ICMI) model encompasses the emergent effects of the following three modeling dimensions:
\begin{enumerate}

\item Real-life temporal interactions, modeled at a \textit{macroscopic} level (daily). Here, we consider the topological structure of the interactions. Who met whom, and when, during each time window. Every day, we assess the likelihood of each node being exposed and infected within a given time window. This probability is the complement of the chance that the node avoids exposure during all of its encounters with infectious nodes in that time window.

\item The \textit{duration} of all interactions: how long each interaction lasted, and was it long enough to result in infection? Interaction duration is modeled at a \textit{microscopic} level and correlates positively with the latent transmitted viral load, which, in turn, is positively correlated with the probability of getting infected ~\cite{{rea2007duration,ferretti2020quantifying}}. 

\item   Individual disease progression modeling: a personal susceptibility parameter mediates the progression and severity of the disease for infected individuals. Following the literature, for COVID-19, this parameter correlates with age~\cite{inde2021age}. Infected individuals, whether symptomatic or asymptomatic~\cite{gandhi2020asymptomatic,buitrago2020occurrence}, become infectious and can transmit the infection to others. Symptomatic individuals are removed from the network once symptoms appear. Recovered individuals cannot be reinfected with the same variant. The individual disease progression encompasses the population diversity, found to critically affect the spread of Covid-19~\cite{aleta2022}.  
\end{enumerate}
\subsubsection*{The ICM model: contagion process without individual outcomes}
Interacting nodes can be in one of the following states: \textit{S}usceptible,  \textit{E}xposed, \textit{I}nfectious, and either \textit{R}ecovered or \textit{R}emoved. All nodes begin in state $S$, but for some initial infectious patients zero nodes in state $I$. Nodes that interact with infectious nodes might become infected, entering state $E$. State $E$ stands for {\em exposed and infected}. Infected nodes become Infectious, thus entering state $I$. Infectious nodes in state $I$ transition to state $R$. 
The transition between states $S$ and $E$ is probabilistic and immediate. The transition between states $E$ and $I$ and from $I$ to $R$ is merely a function of time. 

In the model, the probability of being exposed is calculated at the end of each time window $\tau$ as the complement of the probability of not getting exposed and infected at any of the interactions during that day.
\begin{equation}
    P_i^{\tau}(S \rightarrow E)= 1-\prod^{N_i^\tau}(1- P_{\text{max}})
    \label{eq:exp} 
\end{equation}

Where $N_i^\tau$ is the subset of infected nodes in the time window $\tau$ that interacted with node $i$ during that time window and thus might potentially expose it to the infection, and $P_{\text{max}}$ is the probability of being infected during a maximal exposure.

\subsubsection*{The ICM model: encounters duration heterogeneity}
\label{sec:ICM}
The probability of infection is inversely correlated with distance and decreases dramatically with it, and is correlated with the duration at the exposure distance ~\cite{rea2007duration,minutesai2018airborne,ferretti2020quantifying}. Hence, the likelihood of getting exposed and infected during each interaction with an infectious node is modeled as a Sigmoid function of the duration of the interaction. 

At each encounter with an infectious node in time window $\tau$, there is a probability for node $i$ to get exposed and infected that is calculated as follows. Let $d_{i,k}$ be a non-zero value for the strength of an edge that enters the focal node $i$ from an {\em infected node k}, where $k \in K$. $K$ is the set of infectious nodes that $i$ encounters in time window $\tau$. Here, the strength of an edge, $d_{i,k}$, corresponds to the duration of a meeting between the focal node $i$ and an infectious neighbor node $k \in K$. Thus, the probability of node $i$ becoming exposed and infected during an encounter with an infectious node $k$ is as follows:
\begin{equation}
    \forall k \in K,  P_{i,k}=
    \begin{cases}
    P_{\epsilon} &  d_{i,k} < D_{\text{min}} \\
    \frac{d_{i,k}}{D_{\text{max}}} & D_{\text{min}} \leq d_{i,k} \leq D_{\text{max}}\\
     1 &  d_{i,k} > D_{\text{max}}
    \end{cases}
\label{eq:2}
\end{equation}
The model assigns an insignificant infection probability, $P_{\epsilon}$, to encounters with infectious nodes that are shorter than the minimal time to infect, $D_{min}$. $D_{min}$ is a property of a pathogen. For meetings longer than the minimal time to infect, the probability of infection is linear to the meeting duration, denoted as the strength of the link, $d_{i,k}$. Meetings longer than a maximal value, $D_{max}$, are considered as having the maximal probability of infection.  

As COVID-19 variants, such as Alpha and Delta, are associated with different exposure levels of transmitted viral load~\cite{luo2021infection,teyssou2021delta}, the virality of such pathogens is correlated in the model with the minimum exposure latency, $D_{\text{min}}$.  
 $D_{\text{max}}$ denotes the duration of the exposure for which the probability of infection is maximal. If the interaction is shorter than $D_{\text{min}}$, the duration is set to a minimal probability that reduces the probability of being infected due to this encounter.

At the end of each time window $\tau$,  the probability of a node $i$ becoming exposed and infected (state $E$) is calculated as the complement of the probability of not being exposed in any of the encounters during that time window with infectious nodes, as follows: 

\begin{equation}
    P_i^{\tau}(S \rightarrow E)= 1-\prod_{k}^K(1-P_{i,k}\cdot P_{\text{max}})
    \label{eq:vl2} 
\end{equation}
Where $P_i^{\tau}(S \rightarrow E)$ is the probability of node $i$ in state {\em Susceptible} to transition from state {\em Susceptible} to state {\em Exposed} following the interactions during the time window $\tau$. 

In the case of a gathering, it is probable that node $i$ may encounter more than one infectious node $k$. We consider that in gatherings in which a node interacts with several infectious others, it is exposed to a  higher viral load. The duration of the interaction in a gathering is calculated as the gathering's duration multiplied by the number of infectious nodes participating in the gathering. 
\subsubsection*{ICMI: adding a detailed personal disease progression modeling}
\label{sec:icmi}
Once infected, personal reactions to the infection differ based on age, comorbidity, and other latent features. To trade off accuracy with simplicity, we encompass all of the above in a personal {\em susceptibility} parameter and correlate it with age~\cite{inde2021age,tisminetzky2022age}. Hence, the higher the individual susceptibility, the more likely the probability of becoming symptomatic and severely ill. 
Disease progression is currently set to follow a known timeline~\cite{polak2020systematic,zayet2020natural}. 

We will now explain the individual medical progression model. Once a person is exposed and infected, the disease progression and timeline depend on their {\em personal susceptibility probability}, referred for node $i$ as $s_i$.

\begin{figure}[h!]
\includegraphics[width=0.8\linewidth]{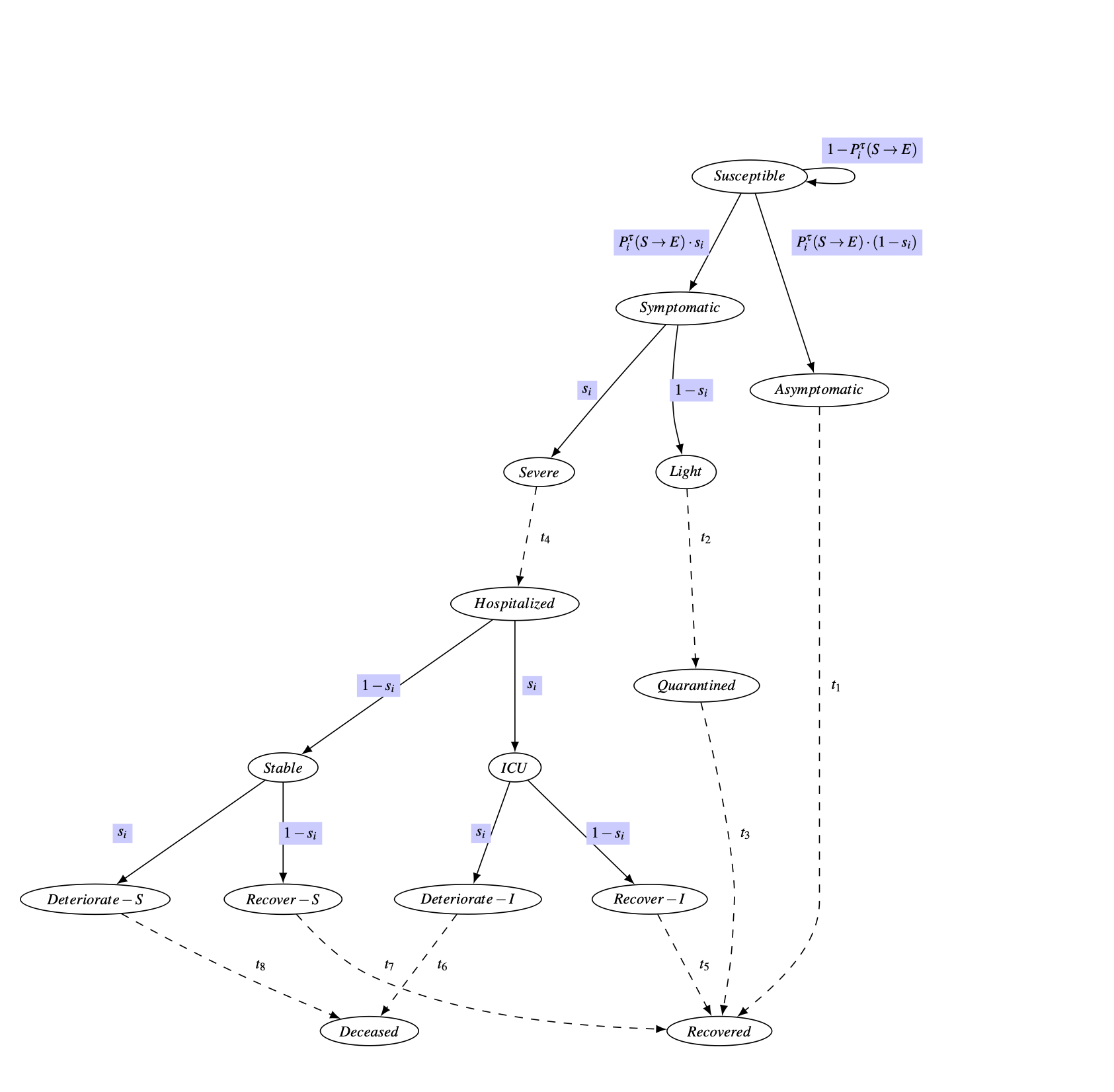} 
\vspace{1cm}
\caption{\textbf{A detailed individual disease progression model for COVID-19}. $s_i$ is the personal susceptibility parameter. Delays ($t_1 \ldots t_8$) are set according to each particular infection's timeline. 
 In the current demonstration, the used values are set according to known COVID-19 timelines available~\cite{polak2020systematic}.}
\label{fig:timeline}
\end{figure}

Figure~\ref{fig:timeline} depicts the model. The initial state is $S$usceptible, and following Eq~\ref{eq:vl2}, a node can transition to states $E$xposed and infected. However, in this state, individuals may be either \textit{Symptomatic} or \textit{Asymptomatic}, depending on their {\em personal susceptibility} parameter, $s_i$. Hence, the transition from state $S$ to $E$ would be to \textit{Symptomatic} with probability $s_i$ and to \textit{Asymptomatic} with probability $1-s_i$.  

The incubation period from exposure to infection differs between asymptomatic and symptomatic people; In the simulation, both are modifiable parameters and rely on found COVID-19 infection parameters~\cite{polak2020systematic}. 

Asymptomatic individuals recover and are placed back into the interactions temporal simulation in the Recovered State, in which they are not susceptible to the disease for the run of the simulation. Symptomatic individuals are removed from the interactions temporal network; however, we continue to model their disease progression. 

Sick individuals have severe symptoms with a probability of $s_i$ and light symptoms with a probability of $1-s_i$. People with light symptoms quarantine until they have recovered. People with severe symptoms are hospitalized in the \textit{ICU} in a severe state with probability $s_i$ or a $S$table state with probability $1-s_i$. In both these states, individuals may deteriorate and die in probability $s_i$ or recover in probability $1-s_i$. Individuals who have recovered are re-introduced into the simulation as 'Recovered'. In this state, they are immune to the disease for the remainder of the simulation.

\subsection*{Real-world contact information}
\label{sec:cns}
An ideal dataset for simulating the viral spread of a disease would be the actual population contact tracing. Manual contact tracing is slow and prone to high delays~\cite{kretzschmar2020impact}. Digital contact tracing can aid in that; however, as seen in many countries, highly inaccurate if it is done by digital tracking. Digital apps are ubiquitously suggested~\cite{yoneki2011epimap,ahmed2020survey}. However, the use of their data entails severe privacy issues~\cite{romanini2020privacy}, and information is sparse unless the app is widely adopted~\cite{barrat2020effect}. 

To overcome these hurdles, we use data from the Copenhagen Networks Study (CNS) \cite{sapiezynski2019interaction}. This data includes over 700 students' contact information for over one month, recorded using Bluetooth sensors in mobile phones provided to the participants. The dataset describes a socio-physical activity at a high, non-aggregated resolution; It is temporal, following over thirty days of interactions of over several hundred people.

The CNS proximity information was registered as a function of the Received Signal Strength Indicator (RSSI). Extracting exact distance information from RSSI data is a difficult task~\cite{Liu2014Face,ng2020covid,vu2010joint}. For infection probability, distance is just one of the dimensions in calculating the chance of infection - the directions of standing, ventilation, and the environment are equally significant parameters ~\cite{li2007role,noakes2006modelling,sze2010review}. As this information was unavailable, we chose to model proximity rather than exact distance and direction. Stronger signals correlate roughly with high proximity. Hence, we mined the CNS network for interactions for which the RSSI $\geq -90$, which is the threshold value.

We  model the social network of interactions $\Gamma$ as a sequence of $T$ consecutive undirected weighted temporal graphs $\{G_\tau \in \Gamma, \tau \in T\}$ where each temporal snapshot graph $G_\tau=(V_\tau,E_\tau)$ denotes the subset of interacting nodes $V_\tau$ during the $\tau_{th}$ temporal window and the weighted edges $E_\tau$ the interactions during this time~\cite{holme2012temporal,sekara2016fundamental}.
Each edge is a  distinct interaction. Edge weight corresponds to the {\em duration} of the interaction used in the model described above. 
We further detect gatherings, as in~\cite{sekara2016fundamental}, as the effect of gatherings on contagious epidemic processes was recently researched and found significant~\cite{ciaperoni2020relevance,Ciaperoni:2020aa}. 

\section*{Author contributions statement}

O.M., Y.S., Y.M., and A.A. designed the experiments; Y.M and A.A. wrote the code and performed all the experiments; All authors analyzed the results; O.M. wrote the paper; All authors reviewed the manuscript.

\section*{Additional information}
\textbf{Code availability:} All code and data used in this research are freely available: \url{https://github.com/ScanLab-ossi/covid-simulation}.\\
\textbf{Competing interests:}
The authors declare no competing interests.


\begin{thebibliography}{10}
\urlstyle{rm}
\expandafter\ifx\csname url\endcsname\relax
  \def\url#1{\texttt{#1}}\fi
\expandafter\ifx\csname urlprefix\endcsname\relax\def\urlprefix{URL }\fi
\expandafter\ifx\csname doiprefix\endcsname\relax\def\doiprefix{DOI: }\fi
\providecommand{\bibinfo}[2]{#2}
\providecommand{\eprint}[2][]{\url{#2}}

\bibitem{dufresne2020r0}
\bibinfo{author}{Hébert-Dufresne, L.}, \bibinfo{author}{Althouse, B.~M.},
  \bibinfo{author}{Scarpino, S.~V.} \& \bibinfo{author}{Allard, A.}
\newblock \bibinfo{title}{Beyond $r_0$: Heterogeneity in secondary infections
  and probabilistic epidemic forecasting} (\bibinfo{year}{2020}).
\newblock \eprint{2002.04004}.

\bibitem{sayama2015introduction}
\bibinfo{author}{Sayama, H.}
\newblock \emph{\bibinfo{title}{Introduction to the modeling and analysis of
  complex systems}} (\bibinfo{publisher}{Open SUNY Textbooks},
  \bibinfo{year}{2015}).

\bibitem{liu2018measurability}
\bibinfo{author}{Liu, Q.-H.} \emph{et~al.}
\newblock \bibinfo{journal}{\bibinfo{title}{Measurability of the epidemic
  reproduction number in data-driven contact networks}}.
\newblock {\emph{\JournalTitle{Proceedings of the National Academy of
  Sciences}}} \textbf{\bibinfo{volume}{115}}, \bibinfo{pages}{12680--12685}
  (\bibinfo{year}{2018}).

\bibitem{inde2021age}
\bibinfo{author}{Inde, Z.} \emph{et~al.}
\newblock \bibinfo{journal}{\bibinfo{title}{Age-dependent regulation of
  sars-cov-2 cell entry genes and cell death programs correlates with covid-19
  severity}}.
\newblock {\emph{\JournalTitle{Science Advances}}}
  \textbf{\bibinfo{volume}{7}}, \bibinfo{pages}{eabf8609}
  (\bibinfo{year}{2021}).

\bibitem{banholzer2022estimating}
\bibinfo{author}{Banholzer, N.}, \bibinfo{author}{Feuerriegel, S.} \&
  \bibinfo{author}{Vach, W.}
\newblock \bibinfo{journal}{\bibinfo{title}{Estimating and explaining
  cross-country variation in the effectiveness of non-pharmaceutical
  interventions during covid-19}}.
\newblock {\emph{\JournalTitle{Scientific reports}}}
  \textbf{\bibinfo{volume}{12}}, \bibinfo{pages}{1--12} (\bibinfo{year}{2022}).

\bibitem{abbey2022interaction}
\bibinfo{author}{Abbey, A.}, \bibinfo{author}{Marmor, Y.},
  \bibinfo{author}{Shahar, Y.} \& \bibinfo{author}{Mokryn, O.}
\newblock \bibinfo{journal}{\bibinfo{title}{Exploring the effects of
  activity-preserving time dilation on the dynamic interplay of airborne
  contagion processes and temporal networks using an interaction-driven
  model}}.
\newblock {\emph{\JournalTitle{arXiv preprint arXiv:2202.11591}}}
  (\bibinfo{year}{2022}).

\bibitem{abbey2022analysis}
\bibinfo{author}{Abbey, A.}, \bibinfo{author}{Shahar, Y.} \&
  \bibinfo{author}{Mokryn, O.}
\newblock \bibinfo{journal}{\bibinfo{title}{Analysis of the competition among
  viral strains using a temporal interaction-driven contagion model}}.
\newblock {\emph{\JournalTitle{Scientific Reports}}}
  \textbf{\bibinfo{volume}{12}}, \bibinfo{pages}{1--10} (\bibinfo{year}{2022}).

\bibitem{holme2012temporal}
\bibinfo{author}{Holme, P.} \& \bibinfo{author}{Saram{\"a}ki, J.}
\newblock \bibinfo{journal}{\bibinfo{title}{Temporal networks}}.
\newblock {\emph{\JournalTitle{Physics reports}}}
  \textbf{\bibinfo{volume}{519}}, \bibinfo{pages}{97--125}
  (\bibinfo{year}{2012}).

\bibitem{ENRIGHT201888}
\bibinfo{author}{Enright, J.} \& \bibinfo{author}{Kao, R.~R.}
\newblock \bibinfo{journal}{\bibinfo{title}{Epidemics on dynamic networks}}.
\newblock {\emph{\JournalTitle{Epidemics}}} \textbf{\bibinfo{volume}{24}},
  \bibinfo{pages}{88--97},
  \doiprefix\url{https://doi.org/10.1016/j.epidem.2018.04.003}
  (\bibinfo{year}{2018}).

\bibitem{holme2021fast}
\bibinfo{author}{Holme, P.}
\newblock \bibinfo{journal}{\bibinfo{title}{Fast and principled simulations of
  the sir model on temporal networks}}.
\newblock {\emph{\JournalTitle{Plos one}}} \textbf{\bibinfo{volume}{16}},
  \bibinfo{pages}{e0246961} (\bibinfo{year}{2021}).

\bibitem{rocha2011simulated}
\bibinfo{author}{Rocha, L.~E.}, \bibinfo{author}{Liljeros, F.} \&
  \bibinfo{author}{Holme, P.}
\newblock \bibinfo{journal}{\bibinfo{title}{Simulated epidemics in an empirical
  spatiotemporal network of 50,185 sexual contacts}}.
\newblock {\emph{\JournalTitle{PLoS computational biology}}}
  \textbf{\bibinfo{volume}{7}}, \bibinfo{pages}{e1001109}
  (\bibinfo{year}{2011}).

\bibitem{scholtes2014causality}
\bibinfo{author}{Scholtes, I.} \emph{et~al.}
\newblock \bibinfo{journal}{\bibinfo{title}{Causality-driven slow-down and
  speed-up of diffusion in non-markovian temporal networks}}.
\newblock {\emph{\JournalTitle{Nature communications}}}
  \textbf{\bibinfo{volume}{5}}, \bibinfo{pages}{1--9} (\bibinfo{year}{2014}).

\bibitem{holme2015information}
\bibinfo{author}{Holme, P.}
\newblock \bibinfo{journal}{\bibinfo{title}{Information content of
  contact-pattern representations and predictability of epidemic outbreaks}}.
\newblock {\emph{\JournalTitle{Scientific reports}}}
  \textbf{\bibinfo{volume}{5}}, \bibinfo{pages}{1--12} (\bibinfo{year}{2015}).

\bibitem{delvenne2015diffusion}
\bibinfo{author}{Delvenne, J.-C.}, \bibinfo{author}{Lambiotte, R.} \&
  \bibinfo{author}{Rocha, L.~E.}
\newblock \bibinfo{journal}{\bibinfo{title}{Diffusion on networked systems is a
  question of time or structure}}.
\newblock {\emph{\JournalTitle{Nature communications}}}
  \textbf{\bibinfo{volume}{6}}, \bibinfo{pages}{1--10} (\bibinfo{year}{2015}).

\bibitem{grossmann2020importance}
\bibinfo{author}{Gro{\ss}mann, G.}, \bibinfo{author}{Backenk{\"o}hler, M.} \&
  \bibinfo{author}{Wolf, V.}
\newblock \bibinfo{title}{Importance of interaction structure and stochasticity
  for epidemic spreading: A covid-19 case study}.
\newblock In \emph{\bibinfo{booktitle}{International Conference on Quantitative
  Evaluation of Systems}}, \bibinfo{pages}{211--229}
  (\bibinfo{organization}{Springer}, \bibinfo{year}{2020}).

\bibitem{masuda2020small}
\bibinfo{author}{Masuda, N.} \& \bibinfo{author}{Holme, P.}
\newblock \bibinfo{journal}{\bibinfo{title}{Small inter-event times govern
  epidemic spreading on networks}}.
\newblock {\emph{\JournalTitle{Physical Review Research}}}
  \textbf{\bibinfo{volume}{2}}, \bibinfo{pages}{023163} (\bibinfo{year}{2020}).

\bibitem{wang2021impact}
\bibinfo{author}{Wang, B.}, \bibinfo{author}{Xie, Z.} \& \bibinfo{author}{Han,
  Y.}
\newblock \bibinfo{journal}{\bibinfo{title}{Impact of individual behavioral
  changes on epidemic spreading in time-varying networks}}.
\newblock {\emph{\JournalTitle{Physical Review E}}}
  \textbf{\bibinfo{volume}{104}}, \bibinfo{pages}{044307}
  (\bibinfo{year}{2021}).

\bibitem{herrmann2020covid}
\bibinfo{author}{Herrmann, H.~A.} \& \bibinfo{author}{Schwartz, J.-M.}
\newblock \bibinfo{journal}{\bibinfo{title}{Why covid-19 models should
  incorporate the network of social interactions}}.
\newblock {\emph{\JournalTitle{Physical Biology}}}
  \textbf{\bibinfo{volume}{17}}, \bibinfo{pages}{065008}
  (\bibinfo{year}{2020}).

\bibitem{stopczynski2014measuring}
\bibinfo{author}{Stopczynski, A.} \emph{et~al.}
\newblock \bibinfo{journal}{\bibinfo{title}{Measuring large-scale social
  networks with high resolution}}.
\newblock {\emph{\JournalTitle{PloS one}}} \textbf{\bibinfo{volume}{9}}
  (\bibinfo{year}{2014}).

\bibitem{Stopczynski:2015aa}
\bibinfo{author}{Stopczynski, A.}, \bibinfo{author}{Sapiezynski, P.},
  \bibinfo{author}{Pentland, A.~S.} \& \bibinfo{author}{Lehmann, S.}
\newblock \bibinfo{journal}{\bibinfo{title}{Temporal fidelity in dynamic social
  networks}}.
\newblock {\emph{\JournalTitle{The European Physical Journal B}}}
  \textbf{\bibinfo{volume}{88}}, \bibinfo{pages}{249},
  \doiprefix\url{10.1140/epjb/e2015-60549-7} (\bibinfo{year}{2015}).

\bibitem{sapiezynski2019interaction}
\bibinfo{author}{Sapiezynski, P.}, \bibinfo{author}{Stopczynski, A.},
  \bibinfo{author}{Lassen, D.~D.} \& \bibinfo{author}{Lehmann, S.}
\newblock \bibinfo{journal}{\bibinfo{title}{Interaction data from the
  copenhagen networks study}}.
\newblock {\emph{\JournalTitle{Scientific Data}}} \textbf{\bibinfo{volume}{6}},
  \bibinfo{pages}{1--10} (\bibinfo{year}{2019}).

\bibitem{vespignani2020modelling}
\bibinfo{author}{Vespignani, A.} \emph{et~al.}
\newblock \bibinfo{journal}{\bibinfo{title}{Modelling covid-19}}.
\newblock {\emph{\JournalTitle{Nature Reviews Physics}}}
  \textbf{\bibinfo{volume}{2}}, \bibinfo{pages}{279--281}
  (\bibinfo{year}{2020}).

\bibitem{thurner2020network}
\bibinfo{author}{Thurner, S.}, \bibinfo{author}{Klimek, P.} \&
  \bibinfo{author}{Hanel, R.}
\newblock \bibinfo{journal}{\bibinfo{title}{A network-based explanation of why
  most covid-19 infection curves are linear}}.
\newblock {\emph{\JournalTitle{Proceedings of the National Academy of
  Sciences}}} \textbf{\bibinfo{volume}{117}}, \bibinfo{pages}{22684--22689}
  (\bibinfo{year}{2020}).

\bibitem{rea2007duration}
\bibinfo{author}{Rea, E.} \emph{et~al.}
\newblock \bibinfo{journal}{\bibinfo{title}{Duration and distance of exposure
  are important predictors of transmission among community contacts of ontario
  sars cases}}.
\newblock {\emph{\JournalTitle{Epidemiology \& Infection}}}
  \textbf{\bibinfo{volume}{135}}, \bibinfo{pages}{914--921}
  (\bibinfo{year}{2007}).

\bibitem{wilber2022model}
\bibinfo{author}{Wilber, M.~Q.} \emph{et~al.}
\newblock \bibinfo{journal}{\bibinfo{title}{A model for leveraging animal
  movement to understand spatio-temporal disease dynamics}}.
\newblock {\emph{\JournalTitle{Ecology Letters}}}
  \textbf{\bibinfo{volume}{25}}, \bibinfo{pages}{1290--1304}
  (\bibinfo{year}{2022}).

\bibitem{smieszek2009mechanistic}
\bibinfo{author}{Smieszek, T.}
\newblock \bibinfo{journal}{\bibinfo{title}{A mechanistic model of infection:
  why duration and intensity of contacts should be included in models of
  disease spread}}.
\newblock {\emph{\JournalTitle{Theoretical Biology and Medical Modelling}}}
  \textbf{\bibinfo{volume}{6}}, \bibinfo{pages}{1--10} (\bibinfo{year}{2009}).

\bibitem{muller2020mobility}
\bibinfo{author}{M{\"u}ller, S.~A.}, \bibinfo{author}{Balmer, M.},
  \bibinfo{author}{Neumann, A.} \& \bibinfo{author}{Nagel, K.}
\newblock \bibinfo{journal}{\bibinfo{title}{Mobility traces and spreading of
  covid-19}}.
\newblock {\emph{\JournalTitle{MedRxiv}}}  (\bibinfo{year}{2020}).

\bibitem{mo2021modeling}
\bibinfo{author}{Mo, B.} \emph{et~al.}
\newblock \bibinfo{journal}{\bibinfo{title}{Modeling epidemic spreading through
  public transit using time-varying encounter network}}.
\newblock {\emph{\JournalTitle{Transportation Research Part C: Emerging
  Technologies}}} \textbf{\bibinfo{volume}{122}}, \bibinfo{pages}{102893}
  (\bibinfo{year}{2021}).

\bibitem{nagel2021realistic}
\bibinfo{author}{Nagel, K.}, \bibinfo{author}{Rakow, C.} \&
  \bibinfo{author}{M{\"u}ller, S.~A.}
\newblock \bibinfo{journal}{\bibinfo{title}{Realistic agent-based simulation of
  infection dynamics and percolation}}.
\newblock {\emph{\JournalTitle{Physica A: Statistical Mechanics and its
  Applications}}} \textbf{\bibinfo{volume}{584}}, \bibinfo{pages}{126322}
  (\bibinfo{year}{2021}).

\bibitem{lorch2020quantifying}
\bibinfo{author}{Lorch, L.} \emph{et~al.}
\newblock \bibinfo{journal}{\bibinfo{title}{Quantifying the effects of contact
  tracing, testing, and containment measures in the presence of infection
  hotspots}}.
\newblock {\emph{\JournalTitle{arXiv preprint arXiv:2004.07641}}}
  (\bibinfo{year}{2020}).

\bibitem{sekara2016fundamental}
\bibinfo{author}{Sekara, V.}, \bibinfo{author}{Stopczynski, A.} \&
  \bibinfo{author}{Lehmann, S.}
\newblock \bibinfo{journal}{\bibinfo{title}{Fundamental structures of dynamic
  social networks}}.
\newblock {\emph{\JournalTitle{Proceedings of the national academy of
  sciences}}} \textbf{\bibinfo{volume}{113}}, \bibinfo{pages}{9977--9982}
  (\bibinfo{year}{2016}).

\bibitem{ciaperoni2020relevance}
\bibinfo{author}{Ciaperoni, M.} \emph{et~al.}
\newblock \bibinfo{journal}{\bibinfo{title}{Relevance of temporal cores for
  epidemic spread in temporal networks}}.
\newblock {\emph{\JournalTitle{Scientific reports}}}
  \textbf{\bibinfo{volume}{10}}, \bibinfo{pages}{1--15} (\bibinfo{year}{2020}).

\bibitem{cencetti2021digital}
\bibinfo{author}{Cencetti, G.} \emph{et~al.}
\newblock \bibinfo{journal}{\bibinfo{title}{Digital proximity tracing on
  empirical contact networks for pandemic control}}.
\newblock {\emph{\JournalTitle{Nature communications}}}
  \textbf{\bibinfo{volume}{12}}, \bibinfo{pages}{1--12} (\bibinfo{year}{2021}).

\bibitem{pung2022using}
\bibinfo{author}{Pung, R.}, \bibinfo{author}{Firth, J.~A.},
  \bibinfo{author}{Spurgin, L.~G.}, \bibinfo{author}{Lee, V.~J.} \&
  \bibinfo{author}{Kucharski, A.~J.}
\newblock \bibinfo{journal}{\bibinfo{title}{Using high-resolution contact
  networks to evaluate sars-cov-2 transmission and control in large-scale
  multi-day events}}.
\newblock {\emph{\JournalTitle{Nature communications}}}
  \textbf{\bibinfo{volume}{13}}, \bibinfo{pages}{1--11} (\bibinfo{year}{2022}).

\bibitem{gandhi2020asymptomatic}
\bibinfo{author}{Gandhi, M.}, \bibinfo{author}{Yokoe, D.~S.} \&
  \bibinfo{author}{Havlir, D.~V.}
\newblock \bibinfo{title}{Asymptomatic transmission, the achilles’ heel of
  current strategies to control covid-19} (\bibinfo{year}{2020}).

\bibitem{shahar2021computing}
\bibinfo{author}{Shahar, Y.} \& \bibinfo{author}{Mokryn, O.}
\newblock \bibinfo{journal}{\bibinfo{title}{A statistical model for early
  estimation of the prevalence and severity of an epidemic from simple tests
  for infection confirmation}}.
\newblock {\emph{\JournalTitle{medRxiv}}}  (\bibinfo{year}{2021}).

\bibitem{PARK2020100392}
\bibinfo{author}{Park, S.~W.}, \bibinfo{author}{Cornforth, D.~M.},
  \bibinfo{author}{Dushoff, J.} \& \bibinfo{author}{Weitz, J.~S.}
\newblock \bibinfo{journal}{\bibinfo{title}{The time scale of asymptomatic
  transmission affects estimates of epidemic potential in the covid-19
  outbreak}}.
\newblock {\emph{\JournalTitle{Epidemics}}} \textbf{\bibinfo{volume}{31}},
  \bibinfo{pages}{100392},
  \doiprefix\url{https://doi.org/10.1016/j.epidem.2020.100392}
  (\bibinfo{year}{2020}).

\bibitem{byambasuren2020estimating}
\bibinfo{author}{Byambasuren, O.} \emph{et~al.}
\newblock \bibinfo{journal}{\bibinfo{title}{Estimating the extent of
  asymptomatic covid-19 and its potential for community transmission:
  systematic review and meta-analysis}}.
\newblock {\emph{\JournalTitle{Official Journal of the Association of Medical
  Microbiology and Infectious Disease Canada}}} \textbf{\bibinfo{volume}{5}},
  \bibinfo{pages}{223--234} (\bibinfo{year}{2020}).

\bibitem{park2020time}
\bibinfo{author}{Park, S.~W.}, \bibinfo{author}{Cornforth, D.~M.},
  \bibinfo{author}{Dushoff, J.} \& \bibinfo{author}{Weitz, J.~S.}
\newblock \bibinfo{journal}{\bibinfo{title}{The time scale of asymptomatic
  transmission affects estimates of epidemic potential in the covid-19
  outbreak}}.
\newblock {\emph{\JournalTitle{Epidemics}}} \textbf{\bibinfo{volume}{31}},
  \bibinfo{pages}{100392} (\bibinfo{year}{2020}).

\bibitem{shahar2023statistical}
\bibinfo{author}{Shahar, Y.} \& \bibinfo{author}{Mokryn, O.}
\newblock \bibinfo{journal}{\bibinfo{title}{A statistical model for early
  estimation of the prevalence and severity of an epidemic or pandemic from
  simple tests for infection confirmation}}.
\newblock {\emph{\JournalTitle{Plos one}}} \textbf{\bibinfo{volume}{18}},
  \bibinfo{pages}{e0280874} (\bibinfo{year}{2023}).

\bibitem{subramanian2021quantifying}
\bibinfo{author}{Subramanian, R.}, \bibinfo{author}{He, Q.} \&
  \bibinfo{author}{Pascual, M.}
\newblock \bibinfo{journal}{\bibinfo{title}{Quantifying asymptomatic infection
  and transmission of covid-19 in new york city using observed cases, serology,
  and testing capacity}}.
\newblock {\emph{\JournalTitle{Proceedings of the National Academy of
  Sciences}}} \textbf{\bibinfo{volume}{118}}, \bibinfo{pages}{e2019716118}
  (\bibinfo{year}{2021}).

\bibitem{lazer2009life}
\bibinfo{author}{Lazer, D.} \emph{et~al.}
\newblock \bibinfo{journal}{\bibinfo{title}{Life in the network: the coming age
  of computational social science}}.
\newblock {\emph{\JournalTitle{Science (New York, NY)}}}
  \textbf{\bibinfo{volume}{323}}, \bibinfo{pages}{721} (\bibinfo{year}{2009}).

\bibitem{mokryn2016role}
\bibinfo{author}{Mokryn, O.}, \bibinfo{author}{Wagner, A.},
  \bibinfo{author}{Blattner, M.}, \bibinfo{author}{Ruppin, E.} \&
  \bibinfo{author}{Shavitt, Y.}
\newblock \bibinfo{journal}{\bibinfo{title}{The role of temporal trends in
  growing networks}}.
\newblock {\emph{\JournalTitle{PloS one}}} \textbf{\bibinfo{volume}{11}},
  \bibinfo{pages}{e0156505} (\bibinfo{year}{2016}).

\bibitem{Peel2015}
\bibinfo{author}{Peel, L.} \& \bibinfo{author}{Clauset, A.}
\newblock \bibinfo{journal}{\bibinfo{title}{{Detecting change points in the
  large-scale structure of evolving networks}}}.
\newblock {\emph{\JournalTitle{29th AAAI Conference on Artificial Intelligence
  (AAAI)}}} \bibinfo{pages}{1--11} (\bibinfo{year}{2015}).
\newblock \eprint{arXiv:1403.0989v1}.

\bibitem{miller2020size}
\bibinfo{author}{Miller, H.} \& \bibinfo{author}{Mokryn, O.}
\newblock \bibinfo{journal}{\bibinfo{title}{Size agnostic change point
  detection framework for evolving networks}}.
\newblock {\emph{\JournalTitle{Plos one}}} \textbf{\bibinfo{volume}{15}},
  \bibinfo{pages}{e0231035} (\bibinfo{year}{2020}).

\bibitem{sun2013understanding}
\bibinfo{author}{Sun, L.}, \bibinfo{author}{Axhausen, K.~W.},
  \bibinfo{author}{Lee, D.-H.} \& \bibinfo{author}{Huang, X.}
\newblock \bibinfo{journal}{\bibinfo{title}{Understanding metropolitan patterns
  of daily encounters}}.
\newblock {\emph{\JournalTitle{Proceedings of the National Academy of
  Sciences}}} \textbf{\bibinfo{volume}{110}}, \bibinfo{pages}{13774--13779}
  (\bibinfo{year}{2013}).

\bibitem{si2020comparative}
\bibinfo{author}{Si, M.} \emph{et~al.}
\newblock \bibinfo{journal}{\bibinfo{title}{A comparative analysis for
  spatio-temporal spreading patterns of emergency news}}.
\newblock {\emph{\JournalTitle{Scientific Reports}}}
  \textbf{\bibinfo{volume}{10}}, \bibinfo{pages}{1--13} (\bibinfo{year}{2020}).

\bibitem{blondel2015survey}
\bibinfo{author}{Blondel, V.~D.}, \bibinfo{author}{Decuyper, A.} \&
  \bibinfo{author}{Krings, G.}
\newblock \bibinfo{journal}{\bibinfo{title}{A survey of results on mobile phone
  datasets analysis}}.
\newblock {\emph{\JournalTitle{EPJ data science}}}
  \textbf{\bibinfo{volume}{4}}, \bibinfo{pages}{10} (\bibinfo{year}{2015}).

\bibitem{cooper2020sir}
\bibinfo{author}{Cooper, I.}, \bibinfo{author}{Mondal, A.} \&
  \bibinfo{author}{Antonopoulos, C.~G.}
\newblock \bibinfo{journal}{\bibinfo{title}{A sir model assumption for the
  spread of covid-19 in different communities}}.
\newblock {\emph{\JournalTitle{Chaos, Solitons \& Fractals}}}
  \textbf{\bibinfo{volume}{139}}, \bibinfo{pages}{110057}
  (\bibinfo{year}{2020}).

\bibitem{kerr2021covasim}
\bibinfo{author}{Kerr, C.~C.} \emph{et~al.}
\newblock \bibinfo{journal}{\bibinfo{title}{Covasim: an agent-based model of
  covid-19 dynamics and interventions}}.
\newblock {\emph{\JournalTitle{PLOS Computational Biology}}}
  \textbf{\bibinfo{volume}{17}}, \bibinfo{pages}{e1009149}
  (\bibinfo{year}{2021}).

\bibitem{truszkowska2021high}
\bibinfo{author}{Truszkowska, A.} \emph{et~al.}
\newblock \bibinfo{journal}{\bibinfo{title}{High-resolution agent-based
  modeling of covid-19 spreading in a small town}}.
\newblock {\emph{\JournalTitle{Advanced theory and simulations}}}
  \textbf{\bibinfo{volume}{4}}, \bibinfo{pages}{2000277}
  (\bibinfo{year}{2021}).

\bibitem{hinch2021openabm}
\bibinfo{author}{Hinch, R.} \emph{et~al.}
\newblock \bibinfo{journal}{\bibinfo{title}{Openabm-covid19—an agent-based
  model for non-pharmaceutical interventions against covid-19 including contact
  tracing}}.
\newblock {\emph{\JournalTitle{PLoS computational biology}}}
  \textbf{\bibinfo{volume}{17}}, \bibinfo{pages}{e1009146}
  (\bibinfo{year}{2021}).

\bibitem{muller2020realistic}
\bibinfo{author}{M{\"u}ller, S.~A.} \emph{et~al.}
\newblock \bibinfo{journal}{\bibinfo{title}{A realistic agent-based simulation
  model for covid-19 based on a traffic simulation and mobile phone data}}.
\newblock {\emph{\JournalTitle{arXiv preprint arXiv:2011.11453}}}
  (\bibinfo{year}{2020}).

\bibitem{tadic2021microscopic}
\bibinfo{author}{Tadi{\'c}, B.} \& \bibinfo{author}{Melnik, R.}
\newblock \bibinfo{journal}{\bibinfo{title}{Microscopic dynamics modeling
  unravels the role of asymptomatic virus carriers in sars-cov-2 epidemics at
  the interplay between biological and social factors}}.
\newblock {\emph{\JournalTitle{Computers in Biology and Medicine}}}
  \textbf{\bibinfo{volume}{133}}, \bibinfo{pages}{104422}
  (\bibinfo{year}{2021}).

\bibitem{grossmann2021heterogeneity}
\bibinfo{author}{Gro{\ss}mann, G.}, \bibinfo{author}{Backenk{\"o}hler, M.} \&
  \bibinfo{author}{Wolf, V.}
\newblock \bibinfo{journal}{\bibinfo{title}{Heterogeneity matters: Contact
  structure and individual variation shape epidemic dynamics}}.
\newblock {\emph{\JournalTitle{Plos one}}} \textbf{\bibinfo{volume}{16}},
  \bibinfo{pages}{e0250050} (\bibinfo{year}{2021}).

\bibitem{gnanvi2021reliability}
\bibinfo{author}{Gnanvi, J.~E.}, \bibinfo{author}{Salako, K.~V.},
  \bibinfo{author}{Kotanmi, G.~B.} \& \bibinfo{author}{Kaka{\"\i}, R.~G.}
\newblock \bibinfo{journal}{\bibinfo{title}{On the reliability of predictions
  on covid-19 dynamics: A systematic and critical review of modelling
  techniques}}.
\newblock {\emph{\JournalTitle{Infectious Disease Modelling}}}
  \textbf{\bibinfo{volume}{6}}, \bibinfo{pages}{258--272}
  (\bibinfo{year}{2021}).

\bibitem{giordano2020modelling}
\bibinfo{author}{Giordano, G.} \emph{et~al.}
\newblock \bibinfo{journal}{\bibinfo{title}{Modelling the covid-19 epidemic and
  implementation of population-wide interventions in italy}}.
\newblock {\emph{\JournalTitle{Nature medicine}}}
  \textbf{\bibinfo{volume}{26}}, \bibinfo{pages}{855--860}
  (\bibinfo{year}{2020}).

\bibitem{peirlinck2020visualizing}
\bibinfo{author}{Peirlinck, M.} \emph{et~al.}
\newblock \bibinfo{journal}{\bibinfo{title}{Visualizing the invisible: {The}
  effect of asymptomatic transmission on the outbreak dynamics of {COVID}-19}}.
\newblock {\emph{\JournalTitle{Computer Methods in Applied Mechanics and
  Engineering}}} \textbf{\bibinfo{volume}{372}}, \bibinfo{pages}{113410},
  \doiprefix\url{10.1016/j.cma.2020.113410} (\bibinfo{year}{2020}).

\bibitem{hebert2020macroscopic}
\bibinfo{author}{H{\'e}bert-Dufresne, L.}, \bibinfo{author}{Scarpino, S.~V.} \&
  \bibinfo{author}{Young, J.-G.}
\newblock \bibinfo{journal}{\bibinfo{title}{Macroscopic patterns of interacting
  contagions are indistinguishable from social reinforcement}}.
\newblock {\emph{\JournalTitle{Nature Physics}}} \textbf{\bibinfo{volume}{16}},
  \bibinfo{pages}{426--431} (\bibinfo{year}{2020}).

\bibitem{HAMBRIDGE2021325}
\bibinfo{author}{Hambridge, H.~L.}, \bibinfo{author}{Kahn, R.} \&
  \bibinfo{author}{Onnela, J.-P.}
\newblock \bibinfo{journal}{\bibinfo{title}{Examining sars-cov-2 interventions
  in residential colleges using an empirical network}}.
\newblock {\emph{\JournalTitle{International Journal of Infectious Diseases}}}
  \textbf{\bibinfo{volume}{113}}, \bibinfo{pages}{325--330},
  \doiprefix\url{https://doi.org/10.1016/j.ijid.2021.10.008}
  (\bibinfo{year}{2021}).

\bibitem{liu2020viral}
\bibinfo{author}{Liu, Y.} \emph{et~al.}
\newblock \bibinfo{journal}{\bibinfo{title}{Viral dynamics in mild and severe
  cases of covid-19}}.
\newblock {\emph{\JournalTitle{The Lancet infectious diseases}}}
  \textbf{\bibinfo{volume}{20}}, \bibinfo{pages}{656--657}
  (\bibinfo{year}{2020}).

\bibitem{ferretti2020quantifying}
\bibinfo{author}{Ferretti, L.} \emph{et~al.}
\newblock \bibinfo{journal}{\bibinfo{title}{Quantifying sars-cov-2 transmission
  suggests epidemic control with digital contact tracing}}.
\newblock {\emph{\JournalTitle{Science}}} \textbf{\bibinfo{volume}{368}}
  (\bibinfo{year}{2020}).

\bibitem{buitrago2020occurrence}
\bibinfo{author}{Buitrago-Garcia, D.} \emph{et~al.}
\newblock \bibinfo{journal}{\bibinfo{title}{Occurrence and transmission
  potential of asymptomatic and presymptomatic sars-cov-2 infections: A living
  systematic review and meta-analysis}}.
\newblock {\emph{\JournalTitle{PLoS medicine}}} \textbf{\bibinfo{volume}{17}},
  \bibinfo{pages}{e1003346} (\bibinfo{year}{2020}).

\bibitem{aleta2022}
\bibinfo{author}{Aleta, A.} \emph{et~al.}
\newblock \bibinfo{journal}{\bibinfo{title}{Quantifying the importance and
  location of sars-cov-2 transmission events in large metropolitan areas}}.
\newblock {\emph{\JournalTitle{Proceedings of the National Academy of
  Sciences}}} \textbf{\bibinfo{volume}{119}}, \bibinfo{pages}{e2112182119},
  \doiprefix\url{10.1073/pnas.2112182119} (\bibinfo{year}{2022}).
\newblock \eprint{https://www.pnas.org/doi/pdf/10.1073/pnas.2112182119}.

\bibitem{minutesai2018airborne}
\bibinfo{author}{Ai, Z.} \& \bibinfo{author}{Melikov, A.~K.}
\newblock \bibinfo{journal}{\bibinfo{title}{Airborne spread of expiratory
  droplet nuclei between the occupants of indoor environments: A review}}.
\newblock {\emph{\JournalTitle{Indoor Air}}} \textbf{\bibinfo{volume}{28}},
  \bibinfo{pages}{500--524} (\bibinfo{year}{2018}).

\bibitem{luo2021infection}
\bibinfo{author}{Luo, C.~H.} \emph{et~al.}
\newblock \bibinfo{journal}{\bibinfo{title}{Infection with the sars-cov-2 delta
  variant is associated with higher infectious virus loads compared to the
  alpha variant in both unvaccinated and vaccinated individuals}}.
\newblock {\emph{\JournalTitle{medRxiv}}}  (\bibinfo{year}{2021}).

\bibitem{teyssou2021delta}
\bibinfo{author}{Teyssou, E.} \emph{et~al.}
\newblock \bibinfo{journal}{\bibinfo{title}{The delta sars-cov-2 variant has a
  higher viral load than the beta and the historical variants in nasopharyngeal
  samples from newly diagnosed covid-19 patients}}.
\newblock {\emph{\JournalTitle{Journal of Infection}}}
  \textbf{\bibinfo{volume}{83}}, \bibinfo{pages}{e1--e3}
  (\bibinfo{year}{2021}).

\bibitem{tisminetzky2022age}
\bibinfo{author}{Tisminetzky, M.} \emph{et~al.}
\newblock \bibinfo{journal}{\bibinfo{title}{Age, multiple chronic conditions,
  and covid-19: a literature review}}.
\newblock {\emph{\JournalTitle{The Journals of Gerontology: Series A}}}
  \textbf{\bibinfo{volume}{77}}, \bibinfo{pages}{872--878}
  (\bibinfo{year}{2022}).

\bibitem{polak2020systematic}
\bibinfo{author}{Polak, S.~B.}, \bibinfo{author}{Van~Gool, I.~C.},
  \bibinfo{author}{Cohen, D.}, \bibinfo{author}{von~der Th{\"u}sen, J.~H.} \&
  \bibinfo{author}{van Paassen, J.}
\newblock \bibinfo{journal}{\bibinfo{title}{A systematic review of pathological
  findings in covid-19: a pathophysiological timeline and possible mechanisms
  of disease progression}}.
\newblock {\emph{\JournalTitle{Modern Pathology}}}
  \textbf{\bibinfo{volume}{33}}, \bibinfo{pages}{2128--2138}
  (\bibinfo{year}{2020}).

\bibitem{zayet2020natural}
\bibinfo{author}{Zayet, S.}, \bibinfo{author}{Gendrin, V.} \&
  \bibinfo{author}{Klopfenstein, T.}
\newblock \bibinfo{journal}{\bibinfo{title}{Natural history of covid-19: back
  to basics}}.
\newblock {\emph{\JournalTitle{New Microbes and New Infections}}}
  \textbf{\bibinfo{volume}{38}}, \bibinfo{pages}{100815}
  (\bibinfo{year}{2020}).

\bibitem{kretzschmar2020impact}
\bibinfo{author}{Kretzschmar, M.~E.} \emph{et~al.}
\newblock \bibinfo{journal}{\bibinfo{title}{Impact of delays on effectiveness
  of contact tracing strategies for covid-19: a modelling study}}.
\newblock {\emph{\JournalTitle{The Lancet Public Health}}}
  \textbf{\bibinfo{volume}{5}}, \bibinfo{pages}{e452--e459}
  (\bibinfo{year}{2020}).

\bibitem{yoneki2011epimap}
\bibinfo{author}{Yoneki, E.} \& \bibinfo{author}{Crowcroft, J.}
\newblock \bibinfo{title}{Epimap: Towards quantifying contact networks and
  modelling the spread of infections in developing countries}.
\newblock In \emph{\bibinfo{booktitle}{Proceedings of the 1st International
  Conference on Wireless Technologies for Humanitarian Relief}},
  \bibinfo{pages}{233--240} (\bibinfo{year}{2011}).

\bibitem{ahmed2020survey}
\bibinfo{author}{Ahmed, N.} \emph{et~al.}
\newblock \bibinfo{journal}{\bibinfo{title}{A survey of covid-19 contact
  tracing apps}}.
\newblock {\emph{\JournalTitle{IEEE Access}}} \textbf{\bibinfo{volume}{8}},
  \bibinfo{pages}{134577--134601} (\bibinfo{year}{2020}).

\bibitem{romanini2020privacy}
\bibinfo{author}{Romanini, D.}, \bibinfo{author}{Lehmann, S.} \&
  \bibinfo{author}{Kivel{\"a}, M.}
\newblock \bibinfo{journal}{\bibinfo{title}{Privacy and uniqueness of
  neighborhoods in social networks}}.
\newblock {\emph{\JournalTitle{arXiv preprint arXiv:2009.09973}}}
  (\bibinfo{year}{2020}).

\bibitem{barrat2020effect}
\bibinfo{author}{Barrat, A.}, \bibinfo{author}{Cattuto, C.},
  \bibinfo{author}{Kivel{\"a}, M.}, \bibinfo{author}{Lehmann, S.} \&
  \bibinfo{author}{Saram{\"a}ki, J.}
\newblock \bibinfo{journal}{\bibinfo{title}{Effect of manual and digital
  contact tracing on covid-19 outbreaks: a study on empirical contact data}}.
\newblock {\emph{\JournalTitle{medRxiv}}}  (\bibinfo{year}{2020}).

\bibitem{Liu2014Face}
\bibinfo{author}{{Liu}, S.}, \bibinfo{author}{{Jiang}, Y.} \&
  \bibinfo{author}{{Striegel}, A.}
\newblock \bibinfo{journal}{\bibinfo{title}{Face-to-face proximity estimation
  using bluetooth on smartphones}}.
\newblock {\emph{\JournalTitle{IEEE Transactions on Mobile Computing}}}
  \textbf{\bibinfo{volume}{13}}, \bibinfo{pages}{811--823},
  \doiprefix\url{10.1109/TMC.2013.44} (\bibinfo{year}{2014}).

\bibitem{ng2020covid}
\bibinfo{author}{Ng, P.~C.}, \bibinfo{author}{Spachos, P.} \&
  \bibinfo{author}{Plataniotis, K.}
\newblock \bibinfo{journal}{\bibinfo{title}{Covid-19 and your smartphone:
  Ble-based smart contact tracing}}.
\newblock {\emph{\JournalTitle{arXiv preprint arXiv:2005.13754}}}
  (\bibinfo{year}{2020}).

\bibitem{vu2010joint}
\bibinfo{author}{Vu, L.}, \bibinfo{author}{Nahrstedt, K.},
  \bibinfo{author}{Retika, S.} \& \bibinfo{author}{Gupta, I.}
\newblock \bibinfo{title}{Joint bluetooth/wifi scanning framework for
  characterizing and leveraging people movement in university campus}.
\newblock In \emph{\bibinfo{booktitle}{Proceedings of the 13th ACM
  International Conference on Modeling, Analysis, and Simulation of Wireless
  and Mobile Systems}}, \bibinfo{pages}{257–265} (\bibinfo{year}{2010}).

\bibitem{li2007role}
\bibinfo{author}{Li, Y.} \emph{et~al.}
\newblock \bibinfo{journal}{\bibinfo{title}{Role of ventilation in airborne
  transmission of infectious agents in the built environment-a
  multidisciplinary systematic review.}}
\newblock {\emph{\JournalTitle{Indoor air}}} \textbf{\bibinfo{volume}{17}},
  \bibinfo{pages}{2--18} (\bibinfo{year}{2007}).

\bibitem{noakes2006modelling}
\bibinfo{author}{Noakes, C.}, \bibinfo{author}{Beggs, C.},
  \bibinfo{author}{Sleigh, P.} \& \bibinfo{author}{Kerr, K.}
\newblock \bibinfo{journal}{\bibinfo{title}{Modelling the transmission of
  airborne infections in enclosed spaces}}.
\newblock {\emph{\JournalTitle{Epidemiology \& Infection}}}
  \textbf{\bibinfo{volume}{134}}, \bibinfo{pages}{1082--1091}
  (\bibinfo{year}{2006}).

\bibitem{sze2010review}
\bibinfo{author}{Sze~To, G.~N.} \& \bibinfo{author}{Chao, C. Y.~H.}
\newblock \bibinfo{journal}{\bibinfo{title}{Review and comparison between the
  wells--riley and dose-response approaches to risk assessment of infectious
  respiratory diseases}}.
\newblock {\emph{\JournalTitle{Indoor air}}} \textbf{\bibinfo{volume}{20}},
  \bibinfo{pages}{2--16} (\bibinfo{year}{2010}).

\bibitem{Ciaperoni:2020aa}
\bibinfo{author}{Ciaperoni, M.} \emph{et~al.}
\newblock \bibinfo{journal}{\bibinfo{title}{Relevance of temporal cores for
  epidemic spread in temporal networks}}.
\newblock {\emph{\JournalTitle{Scientific Reports}}}
  \textbf{\bibinfo{volume}{10}}, \bibinfo{pages}{12529},
  \doiprefix\url{10.1038/s41598-020-69464-3} (\bibinfo{year}{2020}).

\end{thebibliography}

\end{document}